\documentclass[preprintnumbers,article,amsmath,amssymb,floatfix,10pt,prd,twocolumn,
superscriptaddress,nofootinbib]{revtex4-2}
\usepackage{bm}
\usepackage{amsfonts}
\usepackage{latexsym}
\usepackage[latin1]{inputenc}
\usepackage{graphicx}
\usepackage{amsmath}
\usepackage{palatino}
\usepackage{mathpazo}
\usepackage{textcomp}
\usepackage{float}
\usepackage{booktabs}
\usepackage{dcolumn}
\usepackage{ragged2e}
\usepackage{hyperref}
\hypersetup{colorlinks,citecolor=blue}
\hypersetup{colorlinks=true,linkcolor=red,filecolor=magenta,    urlcolor=blue}
\usepackage{amsmath}
\usepackage{xcolor}
\usepackage{orcidlink}
\usepackage{epsfig}
\usepackage{caption}
\usepackage{subcaption}

\usepackage{commath}
\def\jnl@style{\it}
\def\aaref@jnl#1{{\jnl@style#1}}

\def\aaref@jnl#1{{\jnl@style#1}}

\def\aj{\aaref@jnl{AJ}}                   
\def\apj{\aaref@jnl{ApJ}}                 
\def\apjl{\aaref@jnl{ApJ}}                
\def\apjs{\aaref@jnl{ApJS}}               
\def\apss{\aaref@jnl{Ap\&SS}}             
\def\aap{\aaref@jnl{A\&A}}                
\def\aapr{\aaref@jnl{A\&A~Rev.}}          
\def\aaps{\aaref@jnl{A\&AS}}              
\def\mnras{\aaref@jnl{Mon.~Not.~Roy.~Astron.~Soc.}}             
\def\prd{\aaref@jnl{Phys.~Rev.~D}}        
\def\prc{\aaref@jnl{Phys.~Rev.~C}}  
\def\prl{\aaref@jnl{Phys.~Rev.~Lett.}}    
\def\qjras{\aaref@jnl{QJRAS}}             
\def\skytel{\aaref@jnl{S\&T}}             
\def\ssr{\aaref@jnl{Space~Sci.~Rev.}}     
\def\zap{\aaref@jnl{ZAp}}                 
\def\nat{\aaref@jnl{Nature}}              
\def\aplett{\aaref@jnl{Astrophys.~Lett.}} 
\def\apspr{\aaref@jnl{Astrophys.~Space~Phys.~Res.}} 
\def\physrep{\aaref@jnl{Phys.~Rep.}}      
\def\physscr{\aaref@jnl{Phys.~Scr}}       
\def\commat{\aaref@jnl{Comm.~Math.~Phys.}}              
\def\science{\aaref@jnl{Science}}               
\def\cqg{\aaref@jnl{Classical Quant.~Grav.}}            
\def\jpcs{\aaref@jnl{JPCS}}                                     
\def\ijmpd{\aaref@jnl{Int.~J.~Mod.~Phys.~D}}                    
\def\grg{\aaref@jnl{Gen.~Relat.~Gravit.}}               
\def\rpp{\aaref@jnl{Rep.~Prog.~Phys.}}          
\def\npa{\aaref@jnl{Nucl.~Phys.~A}}        
\def\lrr{\aaref@jnl{Living Rev.~Rel.}}                   
\def\jcap{\aaref@jnl{J.~Cosmology Astropart.~Phys.}}    
\def\rmp{\aaref@jnl{Rev.~Mod.~Phys.}}   
\def\epjc{\aaref@jnl{Eur.~Phys.~J.~C}} 
\def\plb{\aaref@jnl{~Phy.~Lett.~B}} 
\def\mpla{\aaref@jnl{Mod.~Phy.~Lett.~A}} 
\def\arxiv{\aaref@jnl{arxiv.org}}

\allowdisplaybreaks[1]

\begin{document}
\color{black}       
%
\title{Bulk viscous cosmological model in  $f(T,\mathcal{T})$ modified gravity}

\author{Raja Solanki\orcidlink{0000-0001-8849-7688}}
\email{rajasolanki8268@gmail.com}
\affiliation{Department of Mathematics, Birla Institute of Technology and
Science-Pilani,\\ Hyderabad Campus, Hyderabad-500078, India.}
\author{Aaqid Bhat\orcidlink{xxxx}}
\email{aaqid555@gmail.com}
\affiliation{Department of Mathematics, Birla Institute of Technology and
Science-Pilani,\\ Hyderabad Campus, Hyderabad-500078, India.}
\author{P.K. Sahoo\orcidlink{0000-0003-2130-8832}}
\email{pksahoo@hyderabad.bits-pilani.ac.in}
\affiliation{Department of Mathematics, Birla Institute of Technology and
Science-Pilani,\\ Hyderabad Campus, Hyderabad-500078, India.}
\date{\today}

\begin{abstract}

This article explores the impact of bulk viscosity on understanding the universe's accelerated expansion within the context of modified $f(T,\mathcal{T})$ gravity, which is an extension of the $f(T)$ gravitational theory, allowing a broad coupling between the energy-momentum scalar $\mathcal{T}$ and the torsion scalar $T$. We consider two $f(T,\mathcal{T})$ functions, specifically $f(T,\mathcal{T})=\alpha T + \beta \mathcal{T}$ and $f(T,\mathcal{T})=\alpha \sqrt{-T} + \beta \mathcal{T}$, where $\alpha$ and $\beta$ are arbitrary constants, along with the fluid part incorporating the coefficient of bulk viscosity $\zeta=\zeta_0 > 0$. We calculate the analytical solutions of the corresponding field equations for a flat FLRW environment, and then we constrain the free parameters of the obtained solution using CC, Pantheon+, and the CC+Pantheon+ samples. We perform the Bayesian statistical analysis to estimate the posterior probability utilizing the likelihood function and the MCMC random sampling technique. Further, to assess the effectiveness of our MCMC analysis, we estimate the corresponding AIC and BIC values, and we find that there is strong evidence supporting the assumed viscous modified gravity models for all three data sets. Also, we find that the linear model precisely mimics the $\Lambda$CDM model. We also investigate the evolutionary behavior of some prominent cosmological parameters. We observe that the effective equation of state parameter for both models predict the accelerating behavior of the cosmic expansion phase. In addition, from the statefinder test, we find that the parameters of the considered MOG models favor the quintessence-type behavior. Further, we observe that the behavior of $Om(z)$ curves corresponding to both models represent a consistent negative slope across the entire domain. We infer that our cosmological setting utilizing $f(T, \mathcal{T})$ gravity models with the viscous matter fluid embodies quintessence-like behavior and can successfully describe the late-time scenario.\\

\textbf{Keywords:} $f(T,\mathcal{T})$ gravity, dark energy, equation of state, and bulk viscosity.

\end{abstract}

\maketitle



\section{Introduction}\label{sec1}
\justifying
 Plenty of cosmological observations, such as the study of Supernovae \cite{R1,R2}, the analysis of Baryon Acoustic Oscillations (BAO) \cite{R3,R4}, the insights gained from the Wilkinson Microwave Anisotropy Probe (WMAP) \cite{R5}, and the study of Cosmic Microwave Background Radiation (CMBR) \cite{R6,R7}, a compelling body of evidence has emerged in support of this cosmic acceleration phenomenon. Central to these findings is the $\Lambda$CDM model, a conceptual framework that incorporates a crucial dark energy component, symbolized as $\Lambda$. However, despite its compatibility with observations, the $\Lambda$CDM model faces well-known issues, such as cosmic coincidence and cosmological constant problem \cite{R8,R9}. One possible solution to circumvent the challenge posed by dark energy is to alter the geometric framework of general relativity (GR) by introducing a more inclusive action that characterizes gravitational interactions \cite{R10}. 

One such modification utilizes the Weitzenb$\ddot{o}$ck connection instead of the typical Levi-Civita connection, characterized by torsion rather than curvature. An intriguing feature of this approach is that the torsion is entirely derived from products of the tetrad's first derivatives, with no involvement of second derivatives in the torsion tensor \cite{R11}. However, torsion does not describe gravitation by geometric representation but rather functions akin to a force \cite{EX1}. In the teleparallel equivalent of general relativity (TEGR), this implies the absence of geodesics, replaced by force equations that resemble the Lorentz force equation in electrodynamics \cite{R12}. Consequently, we can state that gravitational interaction can be alternatively elucidated using curvature through torsion, giving rise to what is known as teleparallel gravity. A modification of TEGR that has garnered attention from cosmologists is known as $f(T)$ gravity, developed as similar to $f(R)$ gravity \cite{R13,R14,CAI}. However, it is important to note that $f(T)$ teleparallel gravity differs from $f(R)$ gravity. One prominent distinction lies in the fact that, for a non-linear $f(R)$ function, gravity manifests as a fourth-order theory, whereas $f(T)$ gravity consistently represents a second-order theory \cite{R15}. Another extension of this theory has been proposed by Harko et al. by incorporating the energy-momentum scalar $\mathcal{T}$ and the torsion scalar $T$, widely known as $f(T,\mathcal{T})$ gravity \cite{R16}. Several interesting applications of the theory have appeared in the literature \cite{R17,HARK,R18,R19,R20,R21,R22}.
 
In earlier studies of the early inflationary period, researchers examined this phenomenon without incorporating the concept of dark energy. In attempting to elucidate this early universe phenomenon, they integrated viscosity within the cosmic fluid to account for dynamic effects. Initially, Eckart made a significant contribution to this subject by formulating a non-causal theory that utilized first-order deviations \cite{C.E.}. In contrast, Israel and Stewart proposed a causal theory, incorporating second-order deviations \cite{W.I.,W.I.-2,W.I.-3}. Notably, the causal theory has been instrumental in examining the observed acceleration of the universe's expansion during later times. When exploring second-order deviations, two viscosity coefficients come into play, namely, shear and bulk viscosity. The velocity gradient corresponds to the shear viscosity vanishes within a homogeneous background spacetime. Consequently, the bulk viscosity coefficient remains significant for a background that is both isotropic and homogeneous, following the FLRW model. Essentially, Einstein's general relativity undergoes geometric modifications that account for the universe's expansion, while the viscosity transport coefficient contributes to the pressure term driving cosmic acceleration. Several interesting insights utilizing viscosity have appeared in the literature \cite{IB-1,IB-2,IB-3,IB-4,IB-5,JM,AVS,MAT}. The present work is organized as follows. In section \eqref{sec2}, we present the geometrical settings of the $f (T,\mathcal{T})$ gravity. In section \eqref{sec3}, we find the analytical solutions of the assumed cosmological $f(T,\mathcal{T})$ models in the presence of cosmic viscous fluid. Further in section \eqref{sec4}, we estimate the corresponding model parameters utilizing the observational data samples through a detailed statistical analysis. Moreover, in section \eqref{sec5}, the evolutionary behavior of some cosmological parameters has been presented. Finally, in section \eqref{sec6}, we highlight the outcomes of our present investigation.

\section{$f (T,\mathcal{T})$ gravity formalism} \label{sec2}
\justifying
In this manuscript, we adopt a notation convention in which Greek indices are employed to encompass the coordinate space-time, while Latin indices are utilized to encompass the tangent space-time. The foundational field of the considered gravity is the vierbein, represented as $e_A(x^\mu)$. At each point $x^\mu$ within the space-time, the vierbein provides an orthonormal basis for the tangent space, satisfying the condition $e_A.e_B = \eta_{AB}$, where $ \eta_{AB} = diag(1,-1,-1,-1)$. Additionally, the vierbein can be represented as the linear combination of the coordinate basis i.e. $e_A=e_A^\mu \partial_\mu$. Consequently, one can have
\begin{equation}\label{2a}
g_{\mu\nu}= \eta_{AB}e^{A}_{\, \mu} e^{B}_{\, \nu}
\end{equation}
Within the framework of teleparallel gravitational theory, an essential concept revolves around the teleparallelism, where the vierbein components at distinct points are parallelized, hence the name teleparallel. In this context, the Weitzenbock connection, denoted as $\Gamma^\alpha_{\nu \mu}= e^\alpha_A \partial_\mu e^A_\nu$, is employed. This choice results in a zero curvature and non-vanishing torsion, in contrast to the Levi-Civita connection, which, when used, yields zero torsion. The corresponding torsion tensor reads as,
\begin{equation}\label{2b}
T^{\alpha}_{\,\mu \nu}= \Gamma^{\alpha}_{\,\nu \mu}- \Gamma^{\,\alpha}_{\mu \nu}= e_{A}^{\,\alpha}\left(\partial_{\mu} e^{A}_{\,\nu}-\partial_{\nu} e^{A}_{\,\mu} \right).
\end{equation}
Further, we define the contortion tensor and the corresponding superpotential tensor as
\begin{eqnarray}
\label{2c}
K^{\mu \nu}_{\,\,\, \alpha} &=& -\frac{1}{2}\left(T^{\mu \nu}_{\,\,\, \alpha} - T^{\nu \mu}_{\,\,\, \alpha} - T_{\alpha}^{\,\, \mu \nu}   \right),\\
\label{2d}
S_{\alpha}^{\,\, \mu \nu} &=& \frac{1}{2}\left(K^{\mu \nu}_{\,\,\, \alpha} + \delta^{\mu}_{\alpha} T^{\lambda \nu}_{\,\,\, \lambda}- \delta^{\nu}_{\alpha} T^{\lambda \mu}_{\,\,\, \lambda}  \right).
\end{eqnarray}
By using equations \eqref{2b} and \eqref{2d}, we define the torsion scalar as 
\begin{equation}\label{2e}
T = S_{\alpha}^{\,\, \mu \nu} T^{\alpha}_{\,\,\mu \nu}= \frac{1}{4} T^{\alpha \mu\nu} T_{\alpha \mu\nu} + \frac{1}{2} T^{\alpha \mu\nu} T_{ \nu\mu \alpha} - T_{\alpha \mu}^{\,\, \,\, \alpha} T^{\nu \mu}_{\,\,\,\, \nu} .
\end{equation}
Consequently, when $T$ is employed in an action and variations are conducted with respect to the vierbeins, the resulting equations align with those of General Relativity. Hence the theory is known as the Teleparallel Equivalent of General Relativity (TEGR). TEGR serves as a foundational framework from which various gravitational modifications can be obtained. One such modification involves $T + f(T)$ in the action, leading to what is known as $f(T)$ gravity. The action for the $f(T,\mathcal{T})$ gravity, that is a further extension to $f(T)$ theory, is given by
\begin{equation}\label{2f}
S =\frac{1}{16\pi G} \int d^4x\, e[T+f(T,\mathcal{T})]+\int d^4x\, e\, \mathcal{L}_m .
\end{equation}
Here, $e=det(e^A_\mu)=\sqrt{-g}$, G is the Newton's constant, and $\mathcal{T}$ represents the trace of the stress-energy tensor $\mathcal{T}^\mu_\nu$. The corresponding field equations acquired by varying the action \eqref{2f} with respect to the vierbeins reads as,
\begin{multline}\label{2g}
(1+f_T)\left[e^{-1}\partial_\mu (e e_{A}^{\,\,\alpha} S_{\alpha}^{\,\, \lambda \mu} )-e_{A}^{\,\, \alpha} T^{\mu}_{\nu \alpha} S_{\mu}^{\,\, \nu \lambda} \right] +  e_{A}^{\,\, \lambda}\left(\frac{f+T}{4}\right)+\\
 \left(f_{TT} \partial_{\mu}T+ f_{T\mathcal{T}}\partial_{\mu}\mathcal{T} \right) e_{A}^{\,\, \alpha} S_{\alpha}^{\,\, \lambda \mu} - f_{\mathcal{T}} \left(\frac{e_{A}^{\,\,\alpha} \mathcal{T}_{\alpha}^{\,\, \lambda} + p e_{A}^{\,\, \lambda}}{2}\right)= \\
 4 \pi G e_{A}^{\,\,\alpha} \mathcal{T}_{\alpha}^{\,\, \lambda}.
\end{multline}
where $f_T={\partial f}/{\partial T} $,$f_\mathcal{T}={\partial f}/{\partial \mathcal{T}} $, $f_{TT}={\partial^2 f}/{\partial T^2} $, and $ f_{T \mathcal{T}}={\partial^2 f}/{\partial T \partial \mathcal{T}}$.\\

Teleparallel gravity and its extensions can be expressed as entirely invariant theories of gravity, preserving both coordinate and local Lorentz symmetries. However, some discussions in the literature suggest that torsion may not qualify as a tensor in teleparallel gravity, implying a potential violation of local Lorentz symmetry and frame dependence in teleparallel gravity theories \cite{GB2}. These perspectives stem from the observation that the teleparallel spin connection adopts a pure-gauge form, allowing for the selection of a specific gauge wherein it nullifies. This particular approach, known as pure tetrad teleparallel gravity, restricts analysis to situations where the connection is nullified, thus lacking covariance. In the TEGR, the teleparallel spin connection does not enter into the field equations governing by the tetrad, and the equations pertaining to the spin connection are naturally satisfied. Consequently, in terms of solving the field equations, the spin connection can be chosen arbitrarily. Notably, this flexibility permits setting the spin connection to zero, leading to the emergence of the purely tetrad formulation of teleparallel gravity.

In the modified TEGR case, the involvement of the spin connection in the action follows a more complex pattern. The variation concerning both the tetrad and the spin connection yields a set of coupled field equations that intricately depend on both variables. However, the resulting equations governing the spin connection equivalent to the antisymmetric component of the equations governing the tetrad. This implies a departure from the liberty to select the spin connection unlike in the case of  ordinary teleparallel gravity. In some symmetric scenarios such as spherically symmetric spacetimes or isotropic spacetime, the field equations for the tetrad exhibit a decoupling between their symmetric and antisymmetric components. Thus, one can easily understand the problem of the pure tetrad formulation in the modified TEGR case and why these theories can easily be misunderstood regarding their local Lorentz invariance. The notion of "good" and "bad" tetrads, initially introduced in the context of $f(T)$ gravity, arises from this understanding \cite{GB1}. Good tetrads, corresponding to a vanishing spin connection, lead to non-trivial solutions of the field equations, contrasting with bad tetrads where $f(T)$ gravity simplifies trivially to ordinary teleparallel gravity. Despite the conceptual and foundational challenges, the pure tetrad formulation, when employing only good tetrads, proves capable of effectively resolving the field equations. Consequently, many findings in the literature, derived through the pure tetrad approach, remain valid.

\section{Cosmological $f(T,\mathcal{T})$ Models with Bulk Viscous Matter}\label{sec3}
\justifying
In order to explore the cosmological implications of the gravity, we assume the following flat FLRW metric,
\begin{equation}\label{3a}
 ds^{2}=dt^{2}-a(t)^{2} \delta_{ij}dx^{i} dx^{j}, 
\end{equation}
where $a(t)$ is the scale factor. The corresponding tetrad field and torsion scalar becomes $e^A_\mu=diag(1,a(t),a(t),a(t))$ and $T=-6H^2$ respectively. Note that the tetrad considered here leads directly to some interesting dynamics means that this tetrad is already in the proper form \cite{MKK} and hence leads to symmetric field equations using the zero connection. However, at the perturbative level one has to consider perturbations to all components of the considered tetrad and solve the antisymmetric part of the field equations, in order to obtain the correct cosmological perturbation theory \cite{AGT}. \\

We explore the influence of the bulk viscosity coefficient, denoted as $\zeta$, in the cosmic evolution. Our assumption involves the viscosity coefficient $\zeta$ adhering to a scaling law, which transforms the Einstein case into a proportional form with respect to the Hubble parameter \cite{IB-1}. Hence, the corresponding energy-momentum tensor is given by
\begin{equation}\label{3b}
\mathcal{T}_{\mu \nu}=(\rho+\bar{p})u_{\mu}u_{\nu} + \bar{p}g_{\mu \nu},
\end{equation}
where $\rho$ is the density of the cosmic matter-energy content and $\bar{p}=p-3\zeta_0 H$ represents the cosmic pressure with viscosity coefficient $\zeta_0$, and $u^\mu=(1,0,0,0)$ is the four velocities vector. \\

The corresponding Friedmann-like equations that describe the gravitational interactions within the background of $f(T,\mathcal{T})$ gravity, when it is filled with a dominating fluid characterized by bulk viscosity, can be expressed as follows \cite{R16}:
\begin{equation}\label{3c}
H^2 =\frac{8\pi G}{3}\rho - \frac{1}{6}\left(f+12H^2f_T \right)+f_\mathcal{T}\left(\frac{\rho+\bar{p}}{3} \right),
\end{equation}
\begin{multline} \label{3d}
\dot{H}= -4\pi G(\rho+\bar{p})-\dot{H}(f_T-12H^2 f_{TT})\\-H(\dot{\rho}-3\dot{\bar{p}}) f_{T \mathcal{T }} - f_\mathcal{T}\left(\frac{\rho+\bar{p}}{2} \right).
\end{multline}
where $\mathcal{T}=\rho-3\bar{p}=\rho+9\zeta_0 H $.\\

By defining the pressure component and energy density of the effective dark energy sector as, 
\begin{equation}\label{ad1}
 \rho_{de}=-\frac{1}{16 \pi G} \left[ f+12f_T H^2-2f_{\mathcal{T}} (\rho+\bar{p}) \right]   
\end{equation}
and
\begin{small}
\begin{equation}\label{ad2}
\begin{split}
p_{de}=  (\rho+\bar{p}) \left[ \frac{1+\frac{f_{\mathcal{T}}}{8\pi G}}{1+f_T-12 H^2 f_{TT}+H \frac{d\rho}{dH}\left(1-3\frac{d\bar{p}}{d\rho} \right) f_{T\mathcal{T}} } -1  \right] \\
+ \frac{1}{16 \pi G} \left[ f+12f_T H^2-2f_{\mathcal{T}} (\rho+\bar{p}) \right]     
\end{split}
\end{equation}    
\end{small}
One can rewrite field equations of the $f(T,\mathcal{T})$ gravity as,
\begin{equation}
H^2=\frac{8\pi G}{3} (\rho+\rho_{de})
\end{equation}
and 
\begin{equation}
\dot{H}=-4\pi G (\rho+\bar{p}+\rho_{de}+p_{de})   
\end{equation}
Note that in the $f(T,\mathcal{T})$ gravity, the effective dark energy is not conserved alone, instead there is an effective coupling between normal matter and dark energy and hence we obtained the conservation equation in a combined form, 
\begin{equation}\label{ad3}
 \dot{\rho}_{de}+\dot{\rho}+3H (\rho+\bar{p}+\rho_{de}+p_{de})=0  
\end{equation}

\subsubsection{Model I}
Through the several examinations of cosmological data and tests within the solar system that validate general relativity, we know that any deviations from the standard general Relativity must be minimal. Consequently, the $f(T)$ function is expected to closely resemble a linear form. Hence, we consider the following linear $f(T,\mathcal{T})$ functional \cite{R23,R24,R25}, 
\begin{equation}\label{3e}
f(T,\mathcal{T})=\alpha T + \beta \mathcal{T}
\end{equation}
where $\alpha$ and $\beta$ are free parameters of the assumed model. Hence, the Friedmann equations \eqref{3c} and \eqref{3d} becomes,
\begin{equation}\label{3f}
 6(1+\alpha)H^2=(\beta+2)\rho - 15\beta \zeta_0 H 
\end{equation}
\begin{equation}\label{3g}
(\beta+2)\dot{H} + 3(\beta+1)H^2 = - \frac{(\beta+1)(1-2\beta)}{(1+\alpha)}\bar{p}
\end{equation}
In, addition, equations \eqref{ad1} and \eqref{ad2} along with the equation \eqref{ad3}, yields,
\begin{equation}\label{cons1}
 \dot{\rho}+ \frac{6(\beta+1)}{(\beta+2) (\alpha+1)}(\rho+\bar{p}) H = \frac{2\dot{H}}{(\beta+2)} \left( 6\alpha H + \frac{15}{2}\beta  \zeta_0  \right)
\end{equation}
By using equations \eqref{3f} and \eqref{3g}, and the transformation $\frac{1}{H}\frac{d}{dt}=\frac{d}{dln(a)}$, we have a following linear differential equation with constant coefficients,
\begin{equation}\label{3h}
\frac{dH}{dln(a)}+\frac{3(\beta+1)}{(\beta+2)}H = \frac{3\zeta_0 (\beta+1)(1-2\beta)}{(1+\alpha)(\beta+2)} 
\end{equation}
On integrating \eqref{3h}, we obtain
\begin{equation}\label{3i}
H(z)=H_0(1+z)^{\frac{3(\beta+1)}{(\beta+2)}}+\frac{\zeta_0(1-2\beta)}{(1+\alpha)}[1-(1+z)^{\frac{3(\beta+1)}{(\beta+2)}}]
\end{equation}
Here $H(0)=H_0$ is present value of the Hubble function. In particular, the choice $\beta=\zeta_0=0$ reduces to a matter-dominated solution of the general relativity.

\subsubsection{Model II}
We consider the following non-linear $f(T,\mathcal{T})$ functional which further leads to a complete non GR analytical solution,
\begin{equation}\label{3e1}
f(T,\mathcal{T})=\alpha \sqrt{-T} + \beta \mathcal{T}
\end{equation}
where $\alpha$ and $\beta$ are free parameters of the assumed model. Hence, the Friedmann equations \eqref{3c} and \eqref{3d} becomes,
\begin{equation}\label{3f1}
 6H^2  + (2\sqrt{6}\alpha + 15\beta \zeta_0 ) H  = (\beta+2)\rho
\end{equation}
\begin{equation}\label{3g1}
 \dot{H}+\frac{3(\beta+1)}{(\beta+2)}H^2=-\frac{3(\beta+1)}{(\beta+2)} \left[\sqrt{\frac{2}{3}}\alpha+\zeta_0(2\beta-1) \right] H
\end{equation}
In, addition, equations \eqref{ad1} and \eqref{ad2} along with the equation \eqref{ad3}, yields,
\begin{equation}\label{cons2}
 \dot{\rho}+ \frac{6(\beta+1)}{(\beta+2)}(\rho+\bar{p}) H = \frac{15}{(\beta+2)} \beta  \zeta_0 \dot{H} 
\end{equation}
By using equations \eqref{3f1} and \eqref{3g1}, and the transformation $\frac{1}{H}\frac{d}{dt}=\frac{d}{dln(a)}$, we have a following linear differential equation with constant coefficients,
\begin{equation}\label{3h1}
 \frac{dH}{dln(a)}+\frac{3(\beta+1)}{(\beta+2)}H=-\frac{3(\beta+1)}{(\beta+2)} \left[\sqrt{\frac{2}{3}}\alpha+\zeta_0(2\beta-1) \right] 
\end{equation}
On integrating \eqref{3h1}, we obtain
\begin{small}
\begin{equation}\label{3i1}
H(z)=H_0(1+z)^{\frac{3(\beta+1)}{(\beta+2)}}+ \left[\sqrt{\frac{2}{3}}\alpha+\zeta_0(2\beta-1) \right]\{(1+z)^{\frac{3(\beta+1)}{(\beta+2)}}-1 \}
\end{equation}    
\end{small}
Here $H(0)=H_0$ is present value of the Hubble function. In particular, the choice $\alpha=\beta=\zeta_0=0$ reduces to a matter-dominated solution of the general relativity.

\section{Estimation of Model Parameters}\label{sec4}
\justifying
In this section, we perform a statistical analysis, comparing the predictions of the considered modified gravity (MOG) models with cosmological observational data to establish limitations on the model's unrestricted variables. We employ Cosmic chronometers sample consisting of $31$ measurements and Pantheon+ sample of $1701$ data points for our analysis.

\subsection{Statistical Methodology}
\justifying
We perform the Bayesian statistical analysis to estimate the posterior probability using the likelihood function and the Markov Chain Monte Carlo (MCMC) random sampling technique. The MCMC is an extensively utilized approach in cosmology to explore the parameter space and produce corresponding probability distributions \cite{R26}. The fundamental concept of MCMC involves generating a Markov chain that samples a model's parameter space based on a given probability distribution. This chain comprises a series of parameter values, wherein each value is derived from the preceding one through a defined set of transition rules tied to a proposal distribution. The proposal distribution provides a fresh parameter value, and its acceptance is dependent upon its posterior probability, considering both the observational data as well as the prior probability function \cite{R27}. After the convergence of the chain, an estimation of the posterior distribution for the parameters can be achieved by calculating the frequency of the parameter values within the chain. Subsequently, the posterior distribution allows for the estimation of optimal parameter values and their associated uncertainties, facilitating predictions for various observables.

\subsection{Cosmic Chronometers}
\justifying
Cosmic chronometers refer to a group of generally elderly, inactive galaxies that have ceased star formation and are distinguished by their spectra and color characteristics \cite{R28}. Data utilized in cosmic chronometers is derived from age measurements of these galaxies at various redshifts. In our study, we employed a compilation of $31$ $H(z)$ data points of passively evolving galaxies in the range of redshift $0.07 \leq z \leq 2.41$, from the Cosmic Chronometers (CC) dataset (listed in the reference \cite{R29}). This $H(z)$ measurements can be achieved using the relation $H(z)=-\frac{1}{1+z} \frac{dz}{dt}$, where $\frac{dz}{dt}$ can be deduce from $\frac{\Delta z}{\Delta t}$. Here, $\Delta z$ signifies the variation in redshift, while $\Delta t$ denotes the variation in age between two galaxies. Numerous recent research investigations have utilized the CC datasets by examining their covariance matrix. This matrix, represented as $C_{ij}$, encompasses a range of factors, such as statistical errors ($C_{ij}^{stat}$), influence from youthful components ($C_{ij}^{young}$), sensitivity to the selected model ($C_{ij}^{model}$), and uncertainties linked to stellar metallicity ($C_{ij}^{stemet}$). The covariance matrix relevant to the CC technique can thus be articulated as follows \cite{COVM},
\begin{equation}\label{4a}
C_{ij}=  C_{ij}^{young} + C_{ij}^{stat} + C_{ij}^{model} + C_{ij}^{stemet}
\end{equation}
In detail, the model covariance, denoted as $C_{ij}^{model}$, can be dissected into various constituents, each signifying a separate origin of uncertainty. These constituents encompass uncertainties arising from the star formation history ($C_{ij}^{SFH}$), the initial mass function ($C_{ij}^{IMF}$), the stellar library ($C_{ij}^{Ste.Lib}$), and the stellar population synthesis model ($C_{ij}^{SPS}$). Consequently, the model covariance is formulated as,
\begin{equation}\label{4a11}
C_{ij}=  C_{ij}^{SFH} + C_{ij}^{IMF} + C_{ij}^{Ste.Lib} + C_{ij}^{SPS}
\end{equation}
To conduct MCMC analysis, it is essential to assess the chi-square function of CCdata, which is defined as follows,
\begin{equation}
\chi^2_{CC} = \Delta A C^{-1} \Delta A^T   
\end{equation}
Here $\Delta A_i = H_{obs.}(z_i)-H_{model}(z_i)$ and $C$ is the covariance matrix whose precise formulation has been discussed in the equation \eqref{4a}.

The best-fit value of parameters of the considered MOG models utilizing the aforementioned CC samples and the prior values as listed in the Table \eqref{Table-1} is presented in the Table \eqref{Table-2}. The corresponding error bar plots for the model comparison are presented in Figs \eqref{f1a} and \eqref{f1b} . Moreover, the corresponding contour plot for the free parameter space $(H_0, \alpha, \beta, \zeta_0)$ within the $1\sigma-3\sigma$ confidence interval is presented in Figs \eqref{f2a} and \eqref{f2b}. \\

\begin{table}[H]
\centering
\begin{tabular}{|c|c|c|}
\hline
Parameters & Gaussian Priors  & Gaussian Priors \\
           &       Model I    &   Model II\\
\hline 
$H_0$ & $[50, 100]$ & $[50, 100]$ \\
$\alpha$ & $[-1,0]$ & $[-10,0]$ \\
$\beta$ & $[0,1]$ & $[-2,2]$ \\
$\zeta_0$ & [0,100] & [0,100]  \\
\hline
\end{tabular}
\caption{MOG model parameters and the corresponding assumed Gaussian priors. }\label{Table-1}
\end{table}

\justify Furthermore, we utilize the $\mathcal{R}^2$ tests to assess the fitting accuracy between the reconstructed outcomes and the observed data. The $\mathcal{R}^2$ value is defined as follows \cite{DPS,Cape-1},
\begin{equation}\label{4b1}
\mathcal{R}^2_{Model I} = 1- \frac{\sum\limits_{k=1}^{31}[H_{obs,k}-H_{th}(z_{k})]^{2}}{\sum\limits_{k=1}^{31}[H_{obs,k}-\bar{H}]^{2}} = 0.901   
\end{equation}
and
\begin{equation}\label{4b2}
\mathcal{R}^2_{Model II} = 1- \frac{\sum\limits_{k=1}^{31}[H_{obs,k}-H_{th}(z_{k})]^{2}}{\sum\limits_{k=1}^{31}[H_{obs,k}-\bar{H}]^{2}} = 0.899   
\end{equation}
where $\bar{H}= \frac{1}{31} \sum\limits_{k=1}^{31}H_{obs,k} $. A higher value of $\mathcal{R}^2$ approaching $1$ indicates a superior degree of fitting. Consequently, the $H(z)$ predicted from the considered MOG models corresponding to best-fit parameter values aligns well with the observed CC data.

\begin{widetext}

\centering
\begin{figure}[H]
{\includegraphics[scale=0.58]{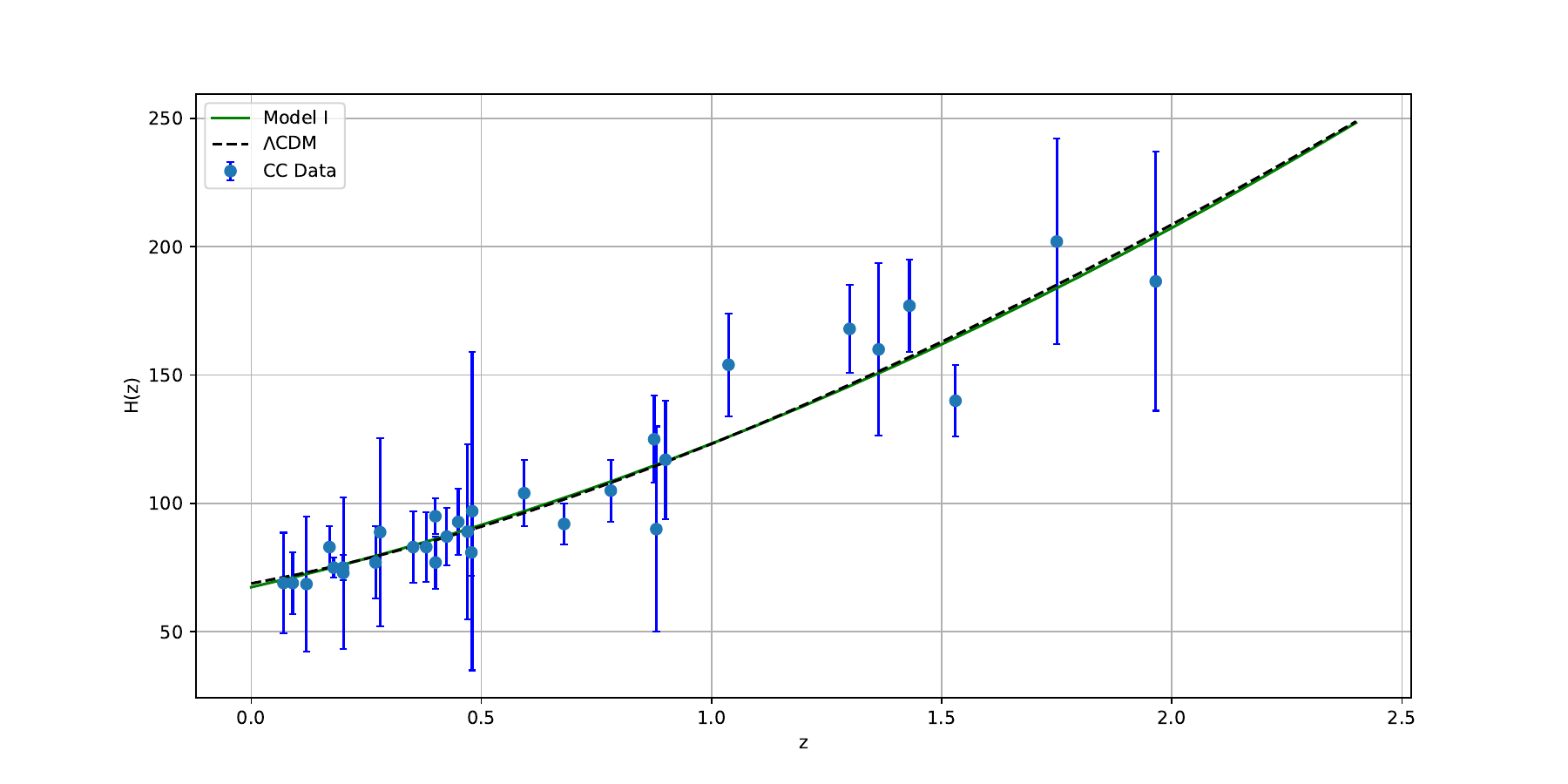}}
{\includegraphics[scale=0.58]{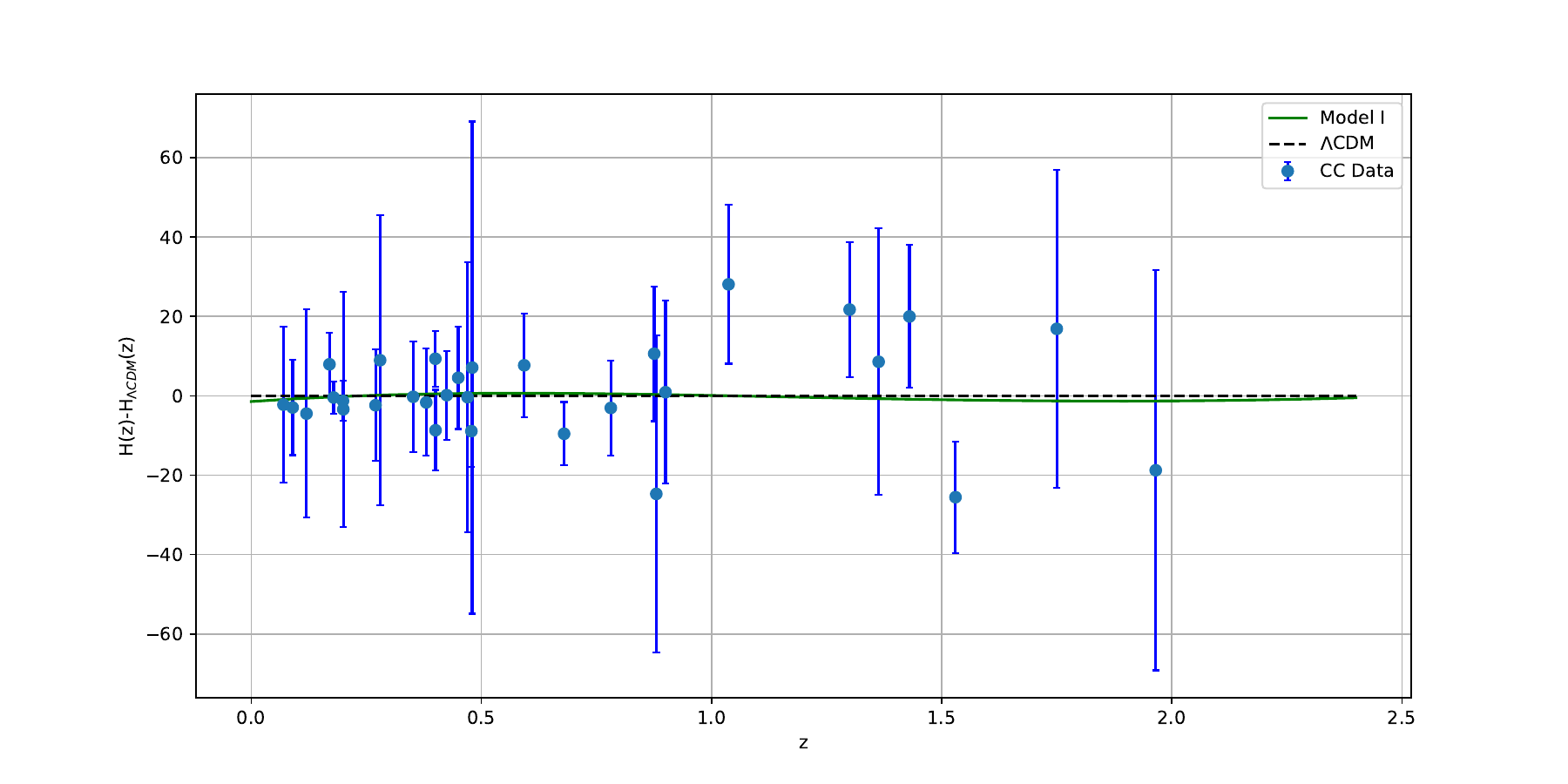}}
\caption{The figure (above) displays the behavior of the Hubble function $H(z)$ for the MOG model I(green curve) and the $\Lambda$CDM model (black dashed curve) along with $31$ CC samples depicted as blue dots, each accompanied by error bars indicated by the blue lines. The figure (below) displays the corresponding variation through the $\Lambda$CDM one. }\label{f1a}
\end{figure}   

\end{widetext}

\begin{widetext}

\centering
\begin{figure}[H]
\includegraphics[scale=0.8]{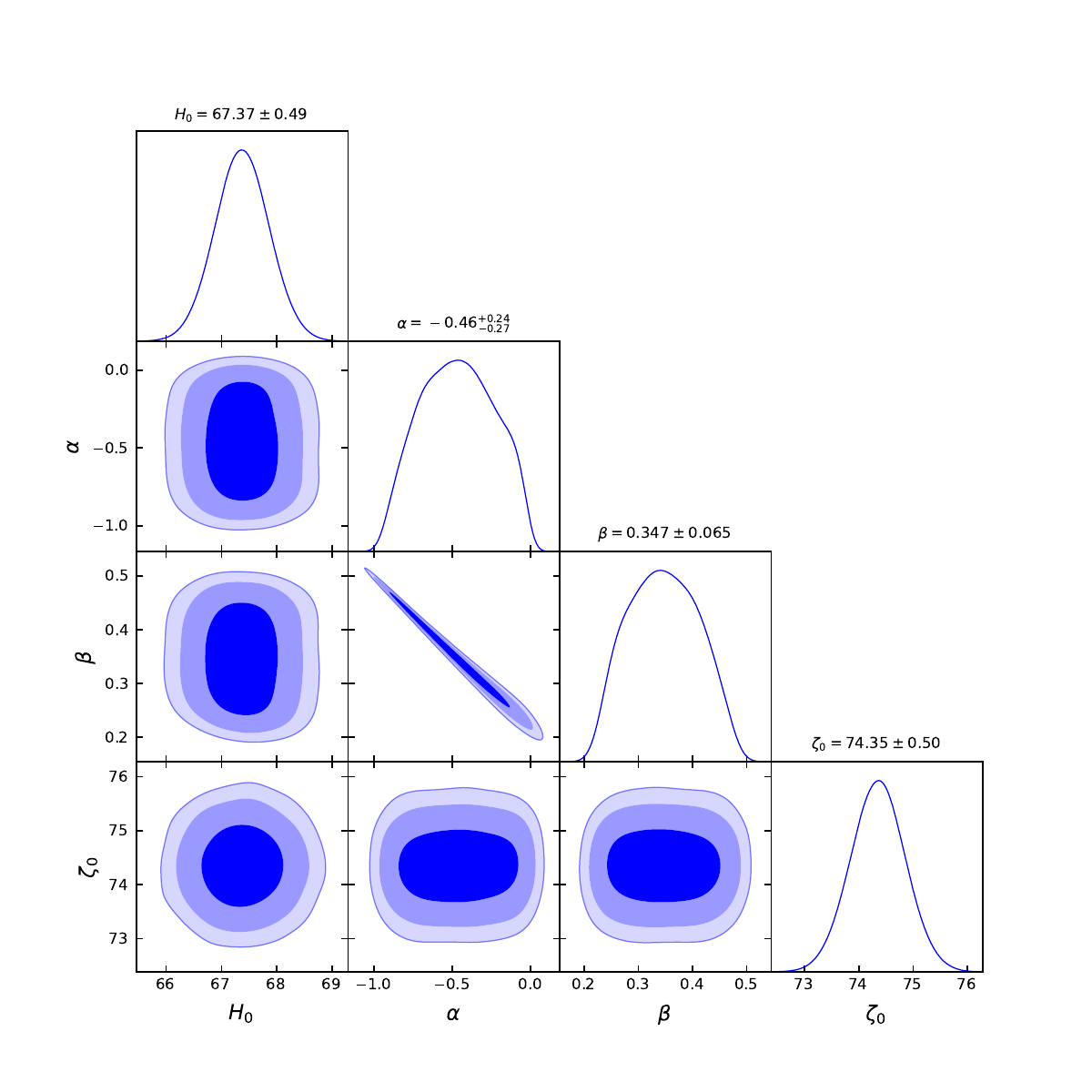}
\caption{The contour plot for the MOG model I corresponding to free parameters $H_0, \alpha, \beta$, and $\zeta_0$ within the $1\sigma-3\sigma$ confidence interval using CC data set.}\label{f2a}
\end{figure}

\end{widetext}  

\begin{widetext}

\centering
\begin{figure}[H]
{\includegraphics[scale=0.58]{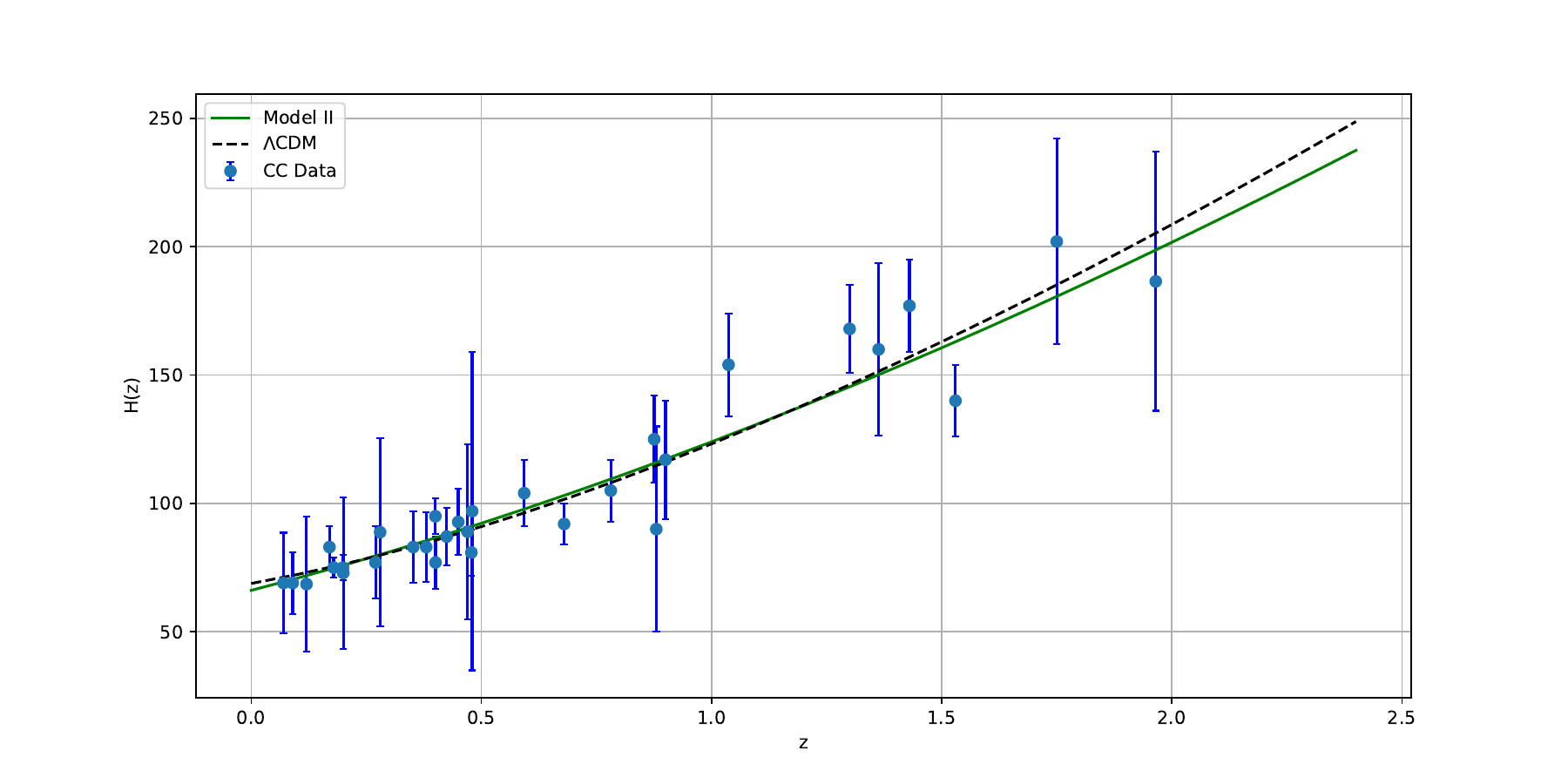}}
{\includegraphics[scale=0.58]{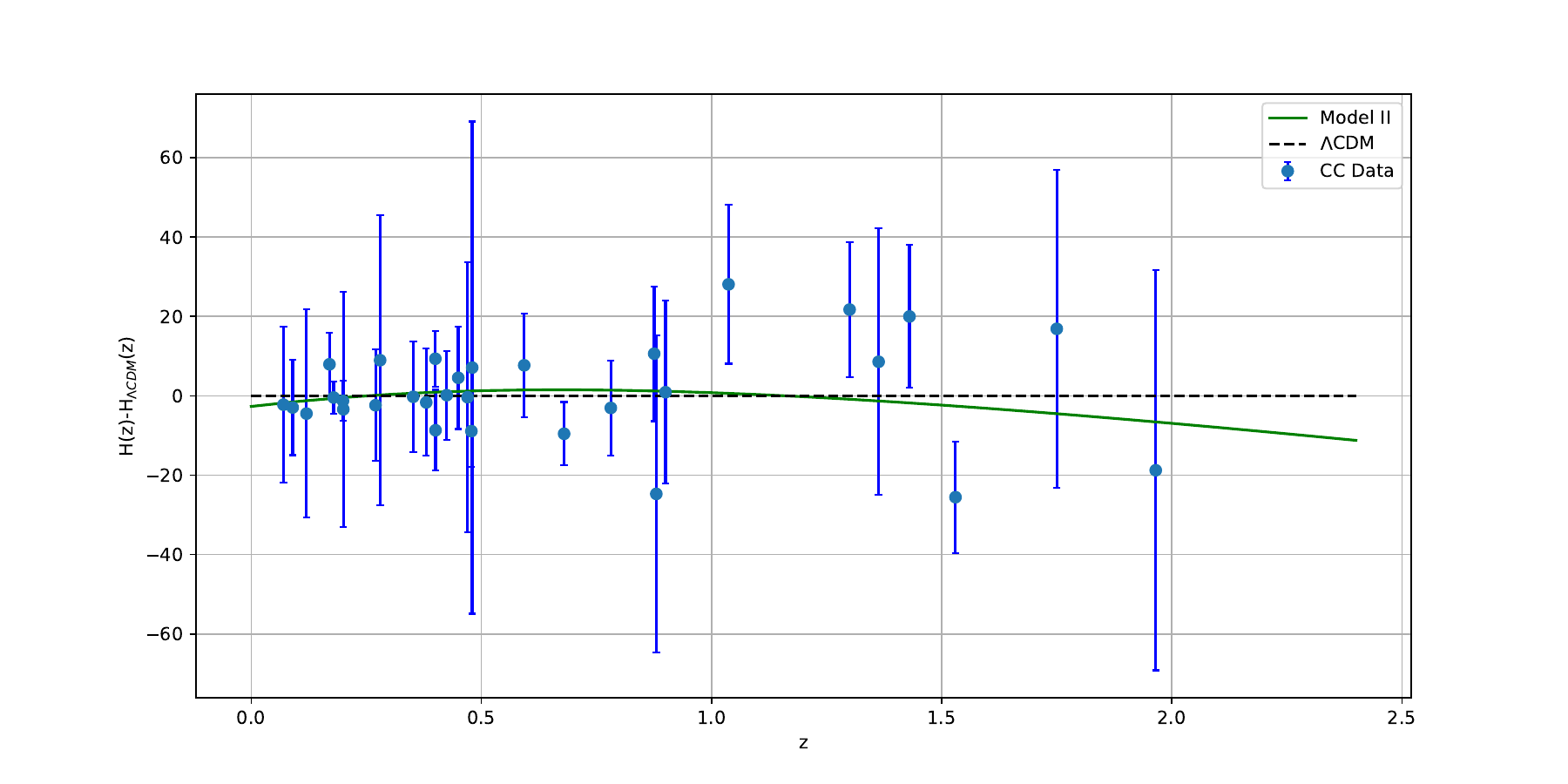}}
\caption{The figure (above) displays the behavior of the Hubble function $H(z)$ for the MOG model II(green curve) and the $\Lambda$CDM model (black dashed curve) along with $31$ CC samples depicted as blue dots, each accompanied by error bars indicated by the blue lines. The figure (below) displays the corresponding variation through the $\Lambda$CDM one. }\label{f1b}
\end{figure}   

\end{widetext}

\begin{widetext}

\centering
\begin{figure}[H]
\includegraphics[scale=0.8]{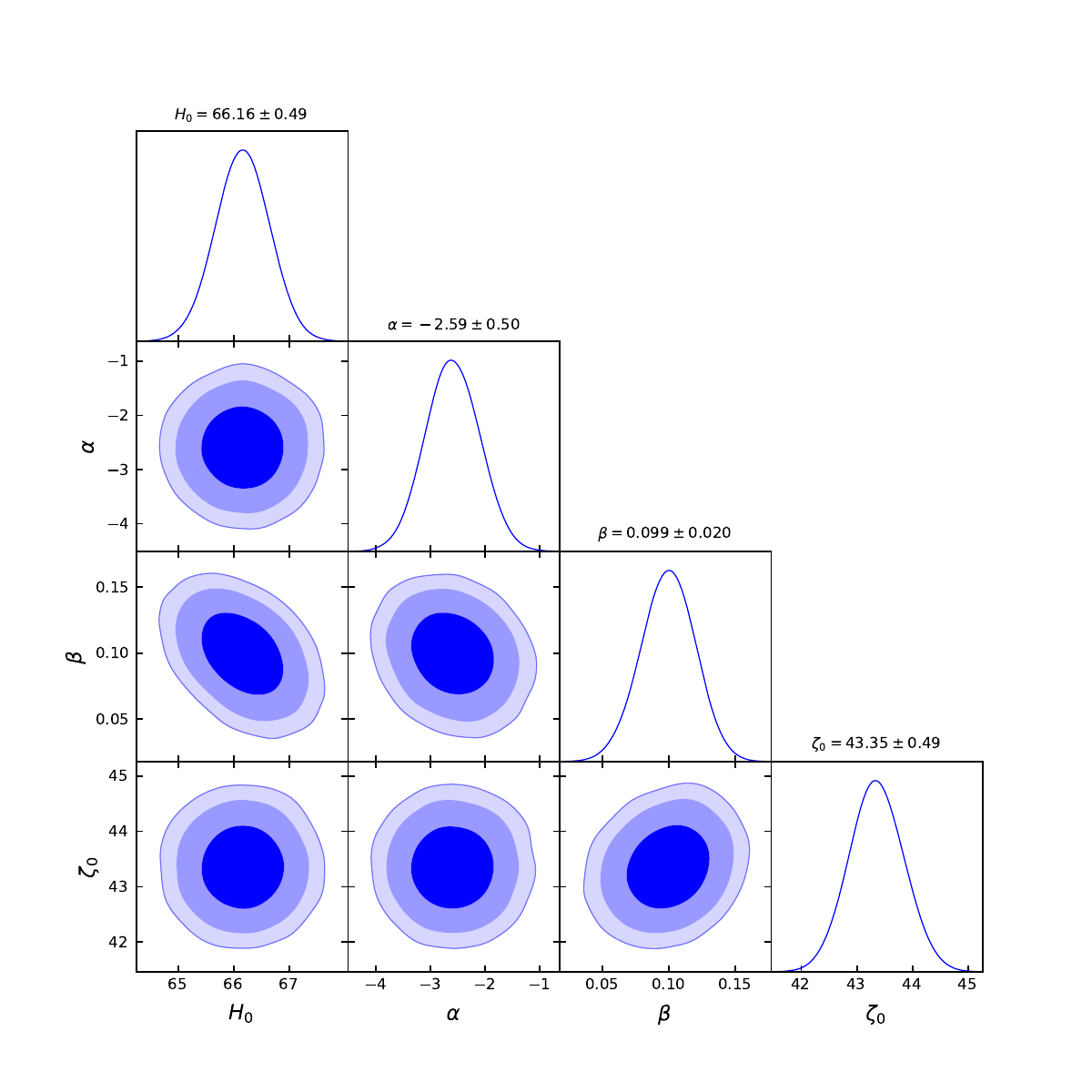}
\caption{The contour plot for the MOG model II corresponding to free parameters $H_0, \alpha, \beta$, and $\zeta_0$ within the $1\sigma-3\sigma$ confidence interval using CC data set.}\label{f2b}
\end{figure}

\end{widetext}

\subsection{Pantheon+}
\justifying
The revised Pantheon+ compilation encompasses an extensive set of $1701$ samples, offering a rich repository of cosmological insights. Encompassing a diverse range of redshifts spanning from $0.001$ to $2.3$, these data points enable researchers to explore the dynamics of cosmic expansion across a considerable span of time. The Pantheon+ samples extend beyond earlier compilations of Type Ia supernovae (SNIa), incorporating the most recent observations. SNIa, known for their brightness, serve as reliable standard candles for estimating relative distances within the cosmos through the distance modulus technique. In recent years, numerous compilations of Type Ia supernova data have emerged, including Union \cite{R30}, Union2 \cite{R31}, Union2.1 \cite{R32}, JLA \cite{R33}, Pantheon \cite{R34}, and the latest addition, Pantheon+ \cite{R35}. The corresponding $\chi^2$ function reads as,
\begin{equation}\label{4c}
\chi^2_{SN}=  D^T C^{-1}_{SN} D,
\end{equation}
Here, $C_{SN}$ denotes the covariance matrix given with the Pantheon+ samples, including both the statistical and systematic uncertainties. Also the vector $D$ is defined as $D=m_{Bi}-M-\mu^{th}(z_i)$, where $m_{Bi}$ and $M$ are respectively the apparent magnitude and the absolute magnitude. Moreover, the expression $\mu^{th}(z_i)$ is the distance module of the theoretical MOG model and can be estimated using the following relation,
\begin{equation}\label{4d}
\mu^{th}(z_i)= 5log_{10} \left[ \frac{D_{L}(z_i)}{1 Mpc}  \right]+25, 
\end{equation}
Here, $D_{L}(z)$ represents the luminosity distance corresponding to the given MOG model and can be estimated as,
\begin{equation}\label{4e}
D_{L}(z)= c(1+z) \int_{0}^{z} \frac{ dx}{H(x,\theta)}
\end{equation}
where, $\theta$ is the usual parameter space.

In contrast to the Pantheon dataset, Pantheon+ compilation successfully resolves the degeneracy between the parameter $H_0$ and $M$ by redefining the vector $D$ as
\begin{equation}\label{4f}
\bar{D} = \begin{cases}
     m_{Bi}-M-\mu_i^{Ceph} & i \in \text{Cepheid hosts} \\
     m_{Bi}-M-\mu^{th}(z_i) & \text{otherwise}
    \end{cases}   
\end{equation}
where $\mu_i^{Ceph}$ independently estimated using Cepheid calibrators. Hence, the relation \eqref{4c} becomes $\chi^2_{SN}=  \bar{D}^T C^{-1}_{SN} \bar{D} $.\\

The best-fit value of parameters of the considered MOG models utilizing the aforementioned Pantheon+ samples and the prior values as listed in the Table \eqref{Table-1} is presented in the Table \eqref{Table-2}. The corresponding error bar plots for the model comparison are presented in Figs \eqref{f3a} and \eqref{f3b}. Moreover, the corresponding contour plot for the free parameter space $(H_0, \alpha, \beta, \zeta_0)$ within the $1\sigma-3\sigma$ confidence interval are presented in Figs \eqref{f4a} and \eqref{f4b}.

\justify Further, we employ the $\mathcal{R}^2$ tests to assess the fitting accuracy between the reconstructed outcomes and the observed data. The $\mathcal{R}^2$ value corresponding to the Pantheon+ samples and estimated best-fit value of parameters is obtained as follows,
\begin{equation}\label{4g1}
\mathcal{R}^2_{Model I} = 1- \frac{\sum\limits_{k=1}^{1701}[\mu_{obs,k}-\mu_{th}(z_{k})]^{2}}{\sum\limits_{k=1}^{1701}[\mu_{obs,k}-\bar{\mu}]^{2}} = 0.996   
\end{equation}
and
\begin{equation}\label{4g2}
\mathcal{R}^2_{Model II} = 1- \frac{\sum\limits_{k=1}^{1701}[\mu_{obs,k}-\mu_{th}(z_{k})]^{2}}{\sum\limits_{k=1}^{1701}[\mu_{obs,k}-\bar{\mu}]^{2}} = 0.996   
\end{equation}
where $\bar{\mu}= \frac{1}{1701} \sum\limits_{k=1}^{1701}\mu_{obs,k} $. It is evident that the $\mu(z)$ predicted from the considered MOG models corresponding to best-fit parameter values aligns well with the observed Pantheon+ samples.

\begin{widetext}

\centering
\begin{figure}[H]
{\includegraphics[scale=0.58]{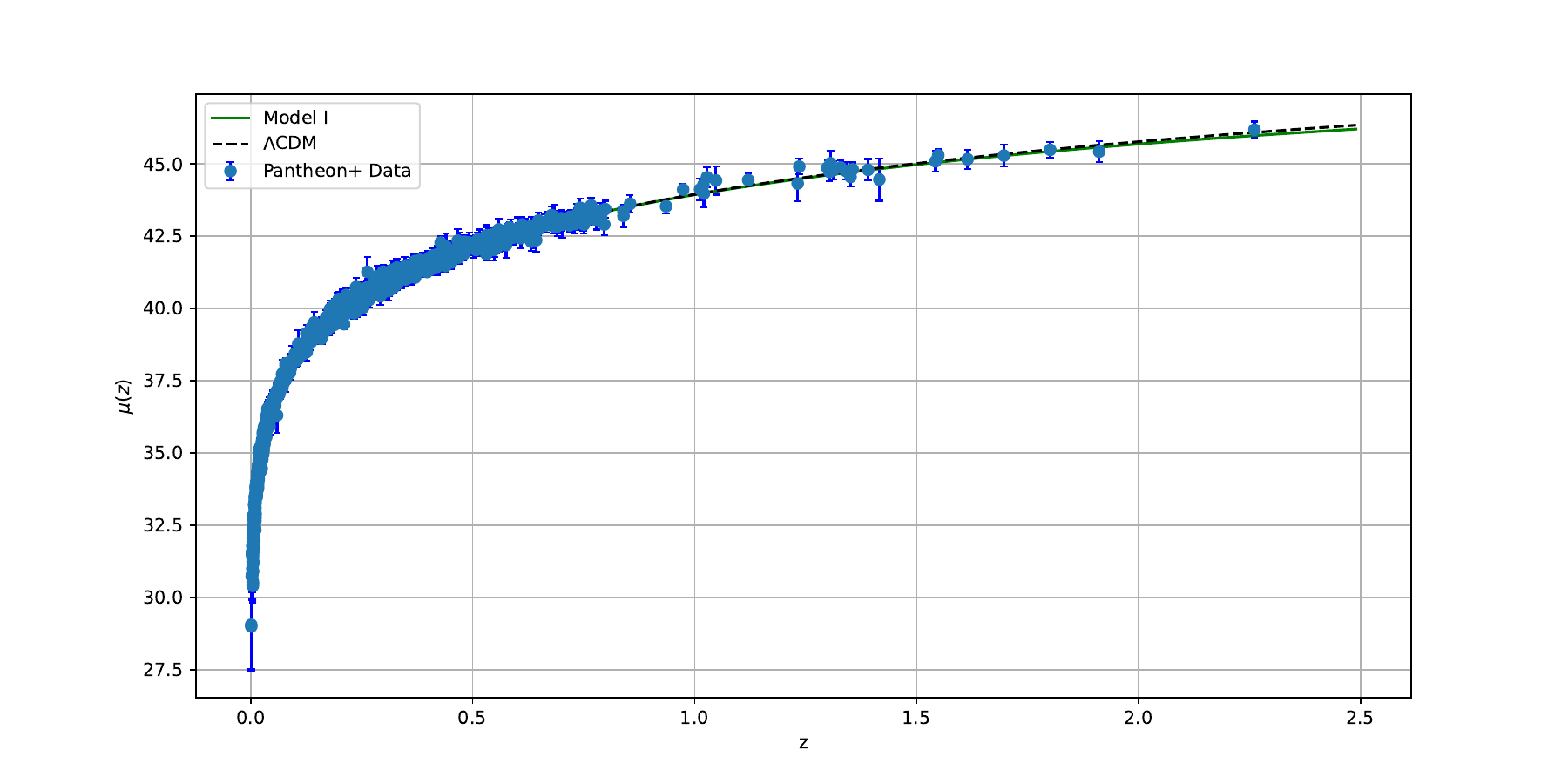}}
{\includegraphics[scale=0.58]{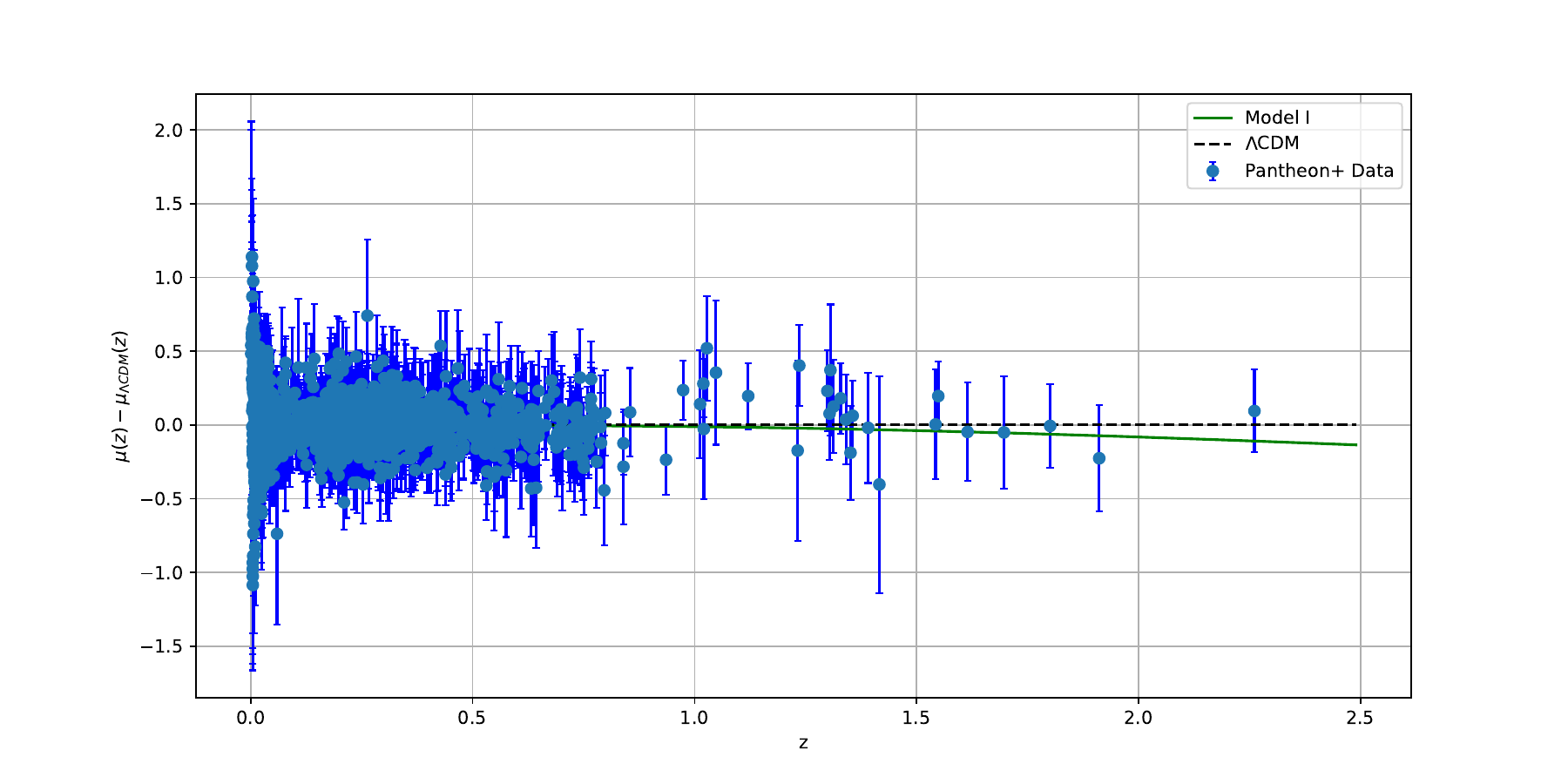}}
\caption{The figure (above) displays the behavior of the distance modulus function $\mu(z)$ for the MOG model I(green curve) and the $\Lambda$CDM model (black dashed curve) along with $1701$ Pantheon+ samples depicted as blue dots, each accompanied by error bars indicated by the blue lines. The figure (below) displays the corresponding variation through the $\Lambda$CDM one.}\label{f3a}
\end{figure}

\end{widetext}

\begin{widetext}

\centering
\begin{figure}[H]
\includegraphics[scale=0.8]{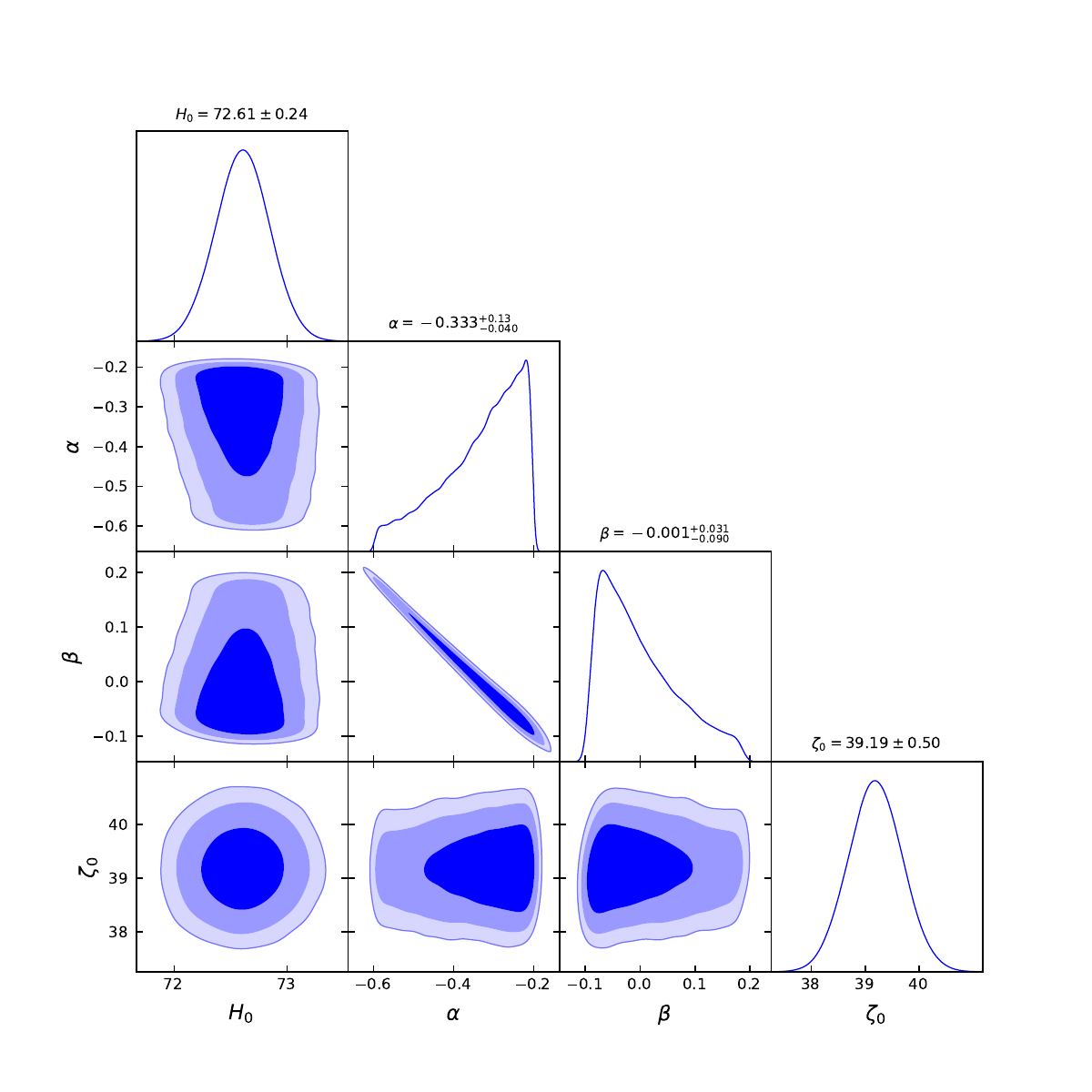}
\caption{The contour plot for the MOG model I corresponding to free parameters $H_0, \alpha, \beta$, and $\zeta_0$ within the $1\sigma-3\sigma$ confidence interval using Pantheon+ data set.}\label{f4a}
\end{figure}

\end{widetext}  

\begin{widetext}

\centering
\begin{figure}[H]
{\includegraphics[scale=0.58]{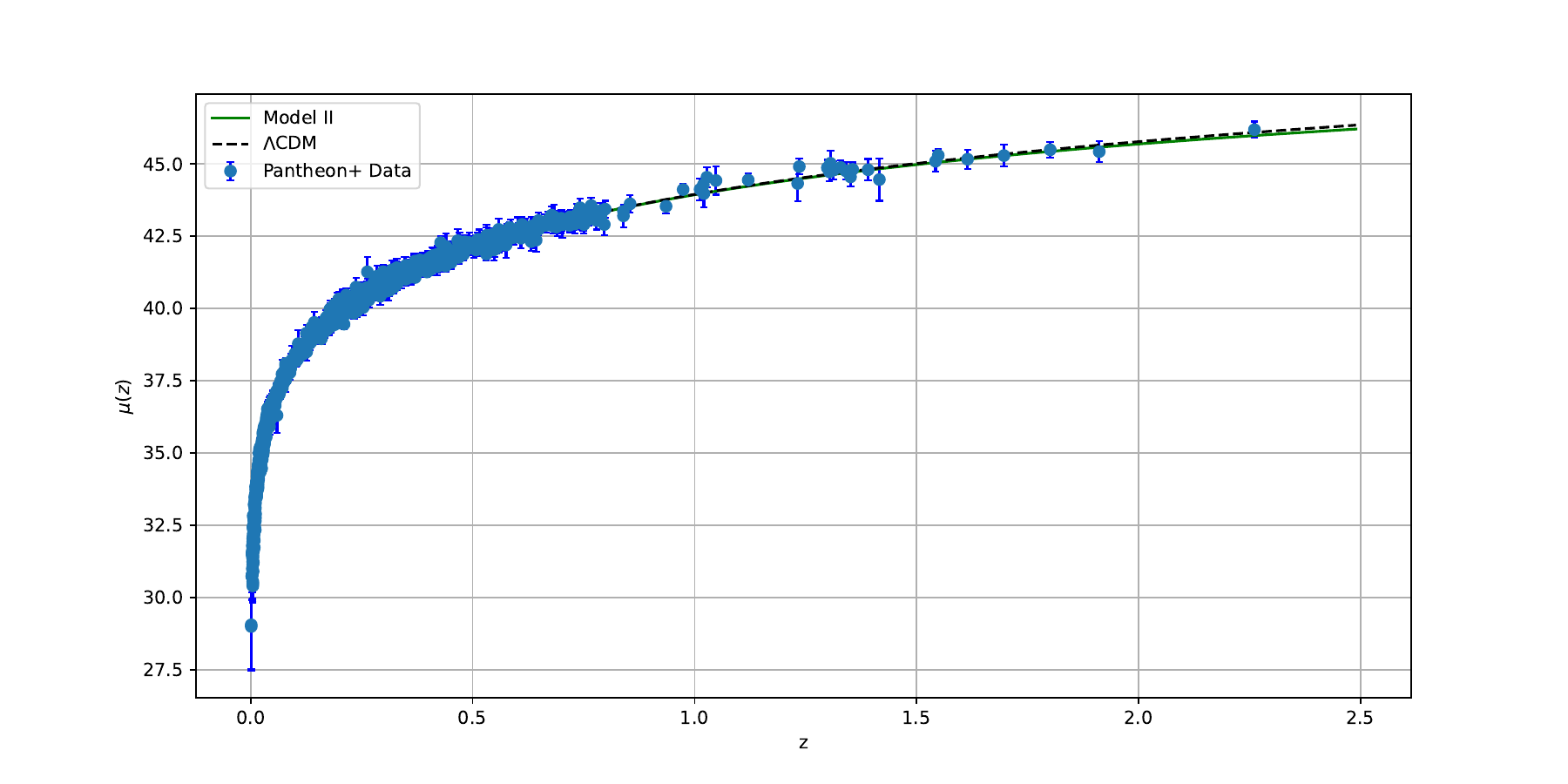}}
{\includegraphics[scale=0.58]{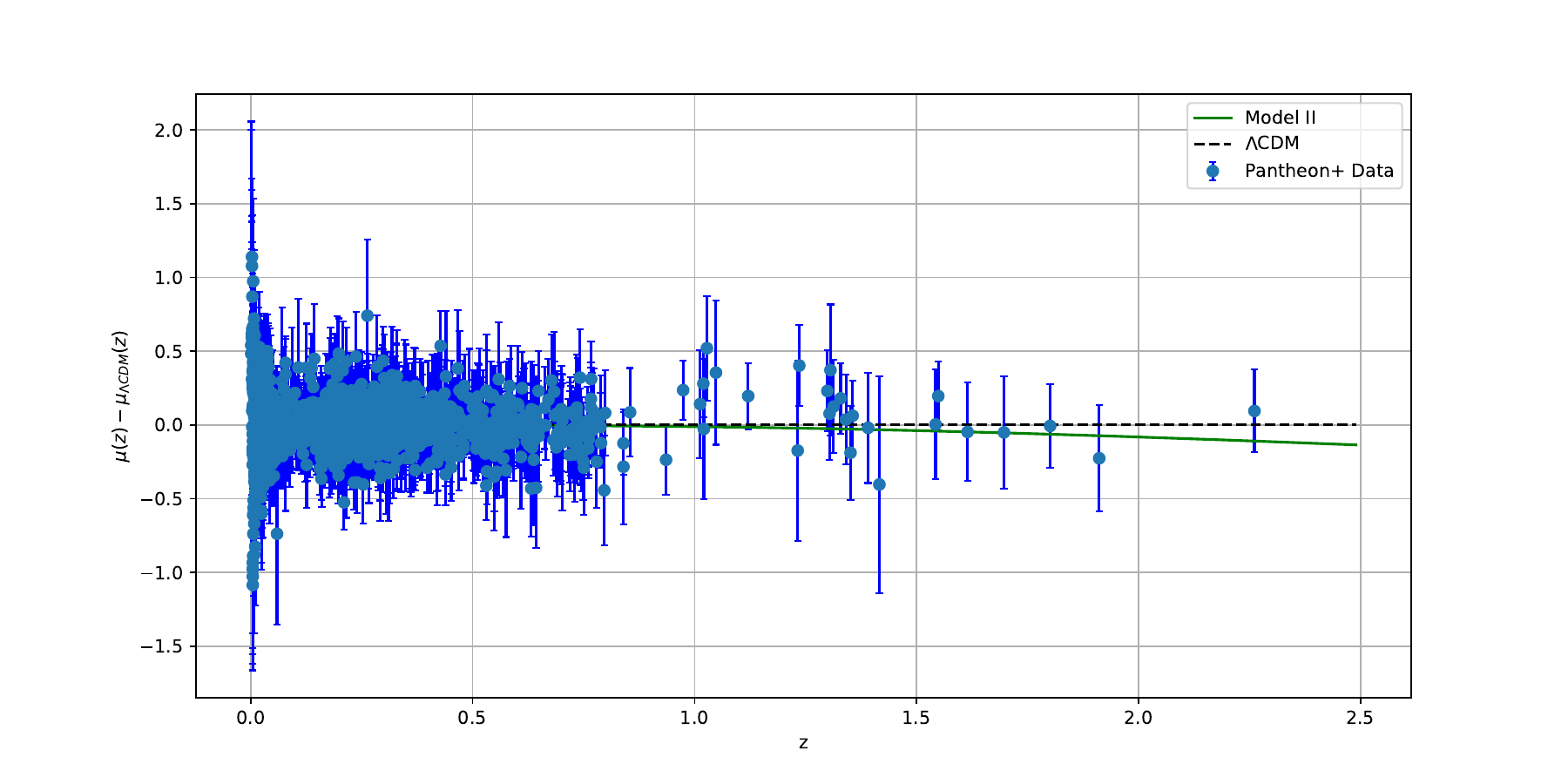}}
\caption{The figure (above) displays the behavior of the distance modulus function $\mu(z)$ for the MOG model II(green curve) and the $\Lambda$CDM model (black dashed curve) along with $1701$ Pantheon+ samples depicted as blue dots, each accompanied by error bars indicated by the blue lines. The figure (below) displays the corresponding variation through the $\Lambda$CDM one.}\label{f3b}
\end{figure}

\end{widetext} 

\begin{widetext}

\centering
\begin{figure}[H]
\includegraphics[scale=0.8]{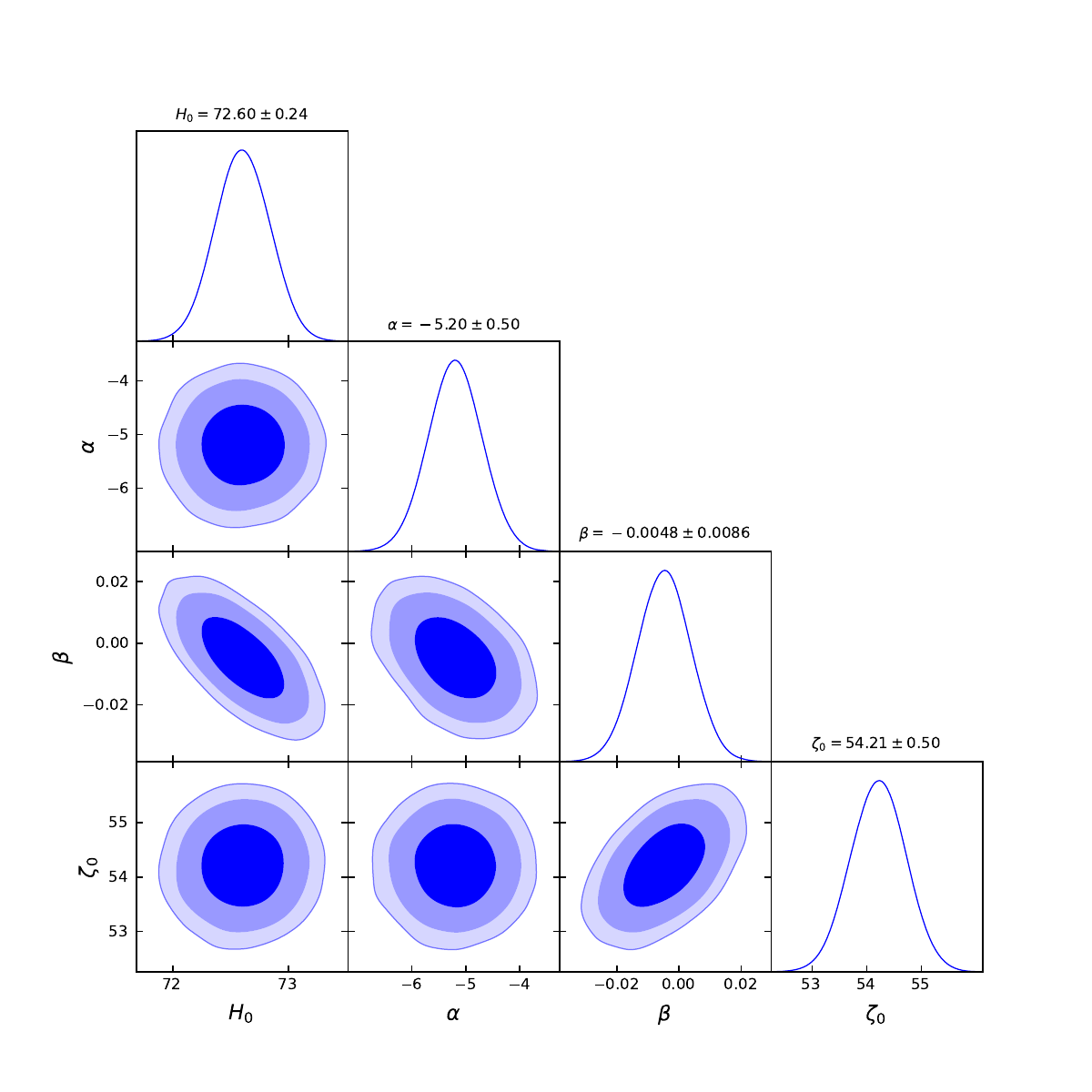}
\caption{The contour plot for the MOG model II corresponding to free parameters $H_0, \alpha, \beta$, and $\zeta_0$ within the $1\sigma-3\sigma$ confidence interval using Pantheon+ data set.}\label{f4b}
\end{figure}

\end{widetext}  

\begin{widetext}

\subsection{CC+Pantheon+}
\justifying
The total $\chi^2$ corresponds to CC+Pantheon+ samples can be acquired by combining the separate contributions as follows,
\begin{equation}\label{4h}
\chi^2_{total}= \chi^2_{CC}+\chi^2_{SN} 
\end{equation}
The best-fit value of parameters of the considered MOG models utilizing the CC+Pantheon+ samples and the prior values as listed in the Table \eqref{Table-1} is presented in the Table \eqref{Table-2}. The corresponding contour plot for the free parameter space $(H_0, \alpha, \beta, \zeta_0)$ within the $1\sigma-3\sigma$ confidence interval are presented in Fig \eqref{f5a} and \eqref{f5b}.

\end{widetext}

\begin{widetext}

\centering
\begin{figure}[H]
\includegraphics[scale=0.8]{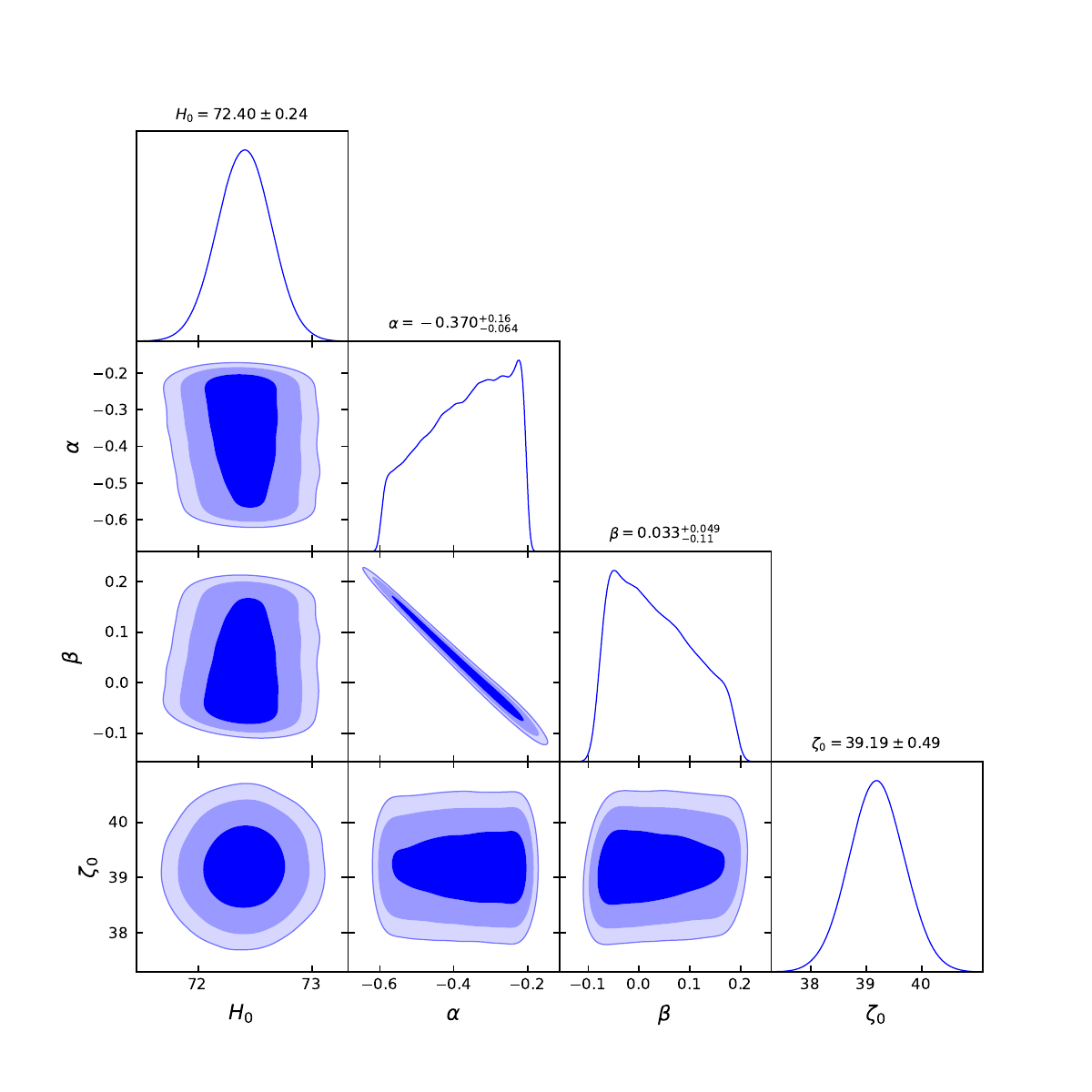}
\caption{The contour plot for the MOG model I corresponding to free parameters $H_0, \alpha, \beta$, and $\zeta_0$ within the $1\sigma-3\sigma$ confidence interval using CC+Pantheon+ data set.}\label{f5a}
\end{figure}

\end{widetext}

\begin{widetext}

\centering
\begin{figure}[H]
\includegraphics[scale=0.8]{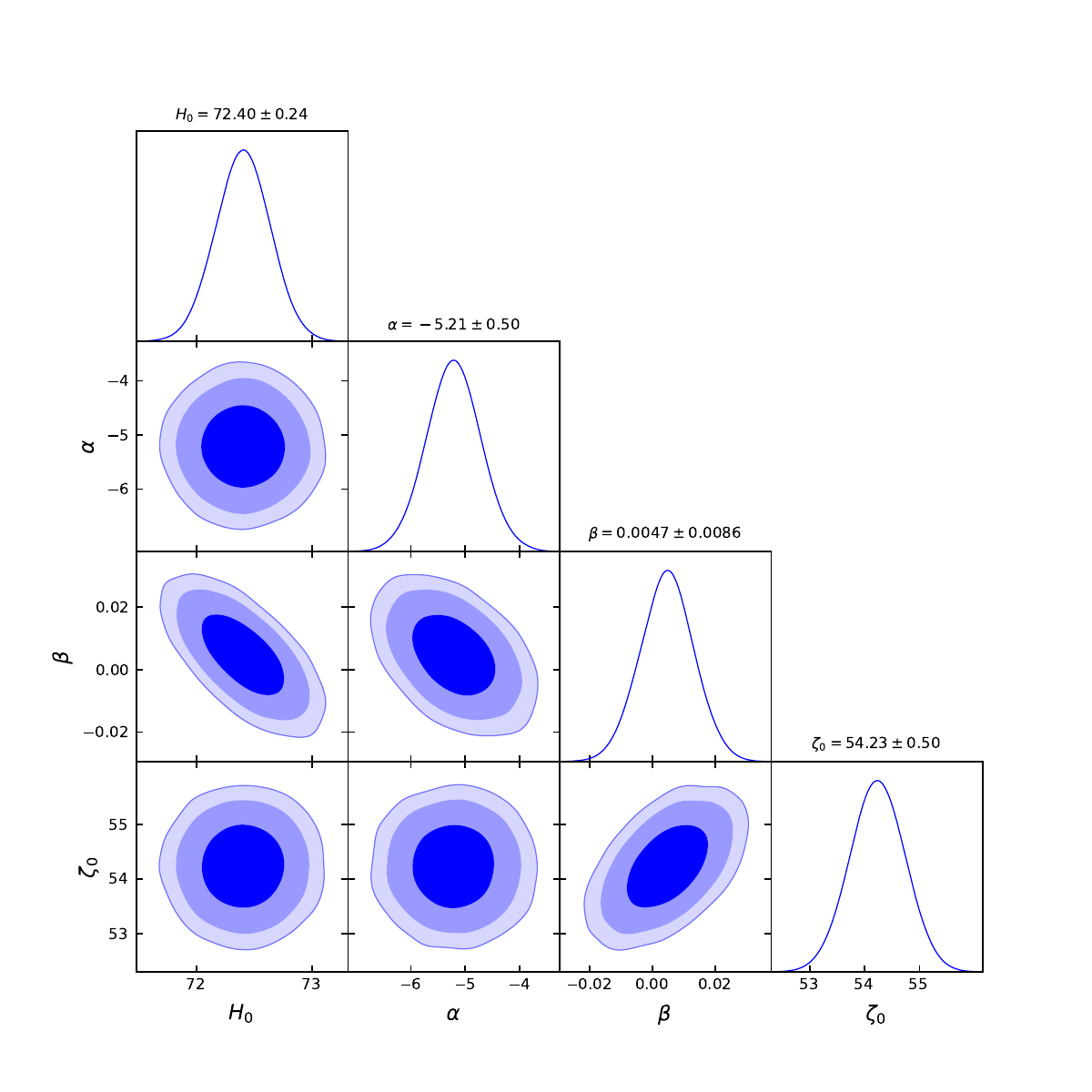}
\caption{The contour plot for the MOG model II corresponding to free parameters $H_0, \alpha, \beta$, and $\zeta_0$ within the $1\sigma-3\sigma$ confidence interval using CC+Pantheon+ data set.}\label{f5b}
\end{figure}

\end{widetext}

\begin{widetext}

\begin{table}[H]
\begin{center}
\begin{tabular}{|c|c|c|c|c|c|}
\hline
& & $H_0$ & $\alpha$  & $\beta$ & $\zeta_0$ \\
\hline 
& CC & $67.37 \pm 0.49$ & $-0.46^{+0.24}_{-0.27} $ & $0.347 \pm 0.065$ & $74.35 \pm 0.50$ \\
Model I & Pantheon+ & $72.61 \pm 0.24$ & $-0.333^{+0.13}_{-0.04} $ & $-0.001^{+0.031}_{-0.090} $ & $39.19 \pm 0.50$ \\
& CC+Pantheon+ & $72.40 \pm 0.24$ &  $-0.37^{+0.16}_{-0.06} $ & $0.033^{+0.049}_{-0.11}$ & $39.19 \pm 0.49$ \\
\hline
& CC & $66.16 \pm 0.49$ & $-2.59 \pm 0.50 $ & $0.099 \pm 0.020$ & $43.35 \pm 0.49$ \\
Model II & Pantheon+ & $72.60 \pm 0.24$  & $-5.20 \pm 0.50 $ & $-0.0048 \pm 0.0086 $ & $54.21 \pm 0.50$ \\
& CC+Pantheon+ & $72.40 \pm 0.24$ & $-5.21 \pm 0.50 $  &  $0.0047 \pm 0.0086 $ & $54.23 \pm 0.50$ \\
\hline
\end{tabular}
\caption{Table shows the best-fit values of the MOG model parameters using different observational data sets.}\label{Table-2}
\end{center}
\end{table}

\end{widetext}

\subsection{Information Criteria}
\justifying
To assess the effectiveness of our MCMC analysis, it is essential to conduct a statistical evaluation utilizing the Akaike Information Criterion (AIC) and Bayesian Information Criterion (BIC) \cite{R36}. The first quantity, AIC, can be defined as follows,
\begin{equation}\label{4i}
AIC = \chi^2_{min} + 2d    
\end{equation}
where $d$ is the number of parameters of the given model. In order to compare our outcomes with the established $\Lambda$CDM model, we define $\Delta AIC = |AIC_{MOG} - AIC_{\Lambda CDM}| $. If $\Delta AIC $ is less than 2, it indicates strong evidence supporting the MOG model, whereas in the range of $ 4 < \Delta AIC \leq 7 $, there is a modest level of evidence favoring the MOG model. Further, if $\Delta AIC$ value exceeds $10$, there is no substantial evidence in favor of the MOG model. The second quantity, BIC, can be defined as follows,
\begin{equation}\label{4j}
BIC = \chi^2_{min} + d ln(N)    
\end{equation}
where $N$ denotes the number of data samples utilized for performing MCMC. Again, if $\Delta BIC $ is less than 2, it indicates strong evidence supporting the MOG model, whereas in the range of $ 2 \leq \Delta BIC < 6 $, there is a modest level of evidence favoring the MOG model. Further, if $\Delta BIC$ value exceeds $6$, there is no substantial evidence in favor of the MOG model. All the relevant values obtained for the considered MOG models are presented in Table \eqref{Table-3}. It is clear from the obtained values there is strong evidence supporting the assumed viscous modified gravity models for all three data sets. In addition, we can conclude that the model I is more precisely mimics the $\Lambda$CDM model.

\begin{widetext}

\begin{table}[H]
\begin{center}
\begin{tabular}{|c|c|c|c|c|c|c|}
\hline
&  &  $\chi^2_{min}$ & AIC & BIC  & $\Delta$AIC & $\Delta$BIC \\
\hline 
&  & MOG \:\:\: $|$\:\:\:  $\Lambda$CDM &  MOG \:\:\: $|$\:\:\:  $\Lambda$CDM &  MOG \:\:\: $|$\:\:\: $\Lambda$CDM &  &\\
\hline
& CC & 32.379 \:\:\:\:\: 32.1322 & 40.379 \:\:\:\:\: 38.1322 & 46.11 \:\:\:\:\: 42.4312 & 2.246 & 3.679\\
Model I & Pantheon+ & 1609.3218 \:\:\:\:\: 1609.9172 & 1617.3218 \:\:\:\:\: 1615.9172 & 1639.0738 \:\:\:\:\: 1632.2312 & 1.404 & 6.842\\
& CC+Pantheon+ & 1641.7008 \:\:\:\:\: 1642.044& 1649.7008 \:\:\:\:\: 1648.0444& 1671.5288 \:\:\:\:\: 1664.204 & 1.651 & 7.108\\
\hline
& CC & 32.568 \:\:\:\:\: 32.1322 & 40.568 \:\:\:\:\: 38.1322 & 46.303 \:\:\:\:\: 42.4312 & 2.435 & 3.872\\
Model II & Pantheon+ & 1610.5494 \:\:\:\:\: 1609.9172 & 1618.5494 \:\:\:\:\: 1615.9172 & 1640.3052 \:\:\:\:\: 1632.2312 & 2.632 & 8.074\\
& CC+Pantheon+ & 1643.1174 \:\:\:\:\: 1642.044& 1651.1174 \:\:\:\:\: 1648.0444& 1672.9455 \:\:\:\:\: 1664.204 & 3.073 & 8.741\\
\hline
\end{tabular}
\caption{Table shows the minimum $\chi^2$ values and the corresponding AIC and BIC values for both considered MOG models.}\label{Table-3}
\end{center}
\end{table}

\end{widetext}

\section{Evolutionary Behavior of Cosmological Parameters}\label{sec5}
\justifying
We present the evolutionary behavior of some prominent cosmological parameters such as effective equation of state (EoS), statefinder, and Om diagnostic parameter, corresponding to the restriction on the model parameters utilizing different observations.

\begin{figure}[H]
\includegraphics[scale=0.47]{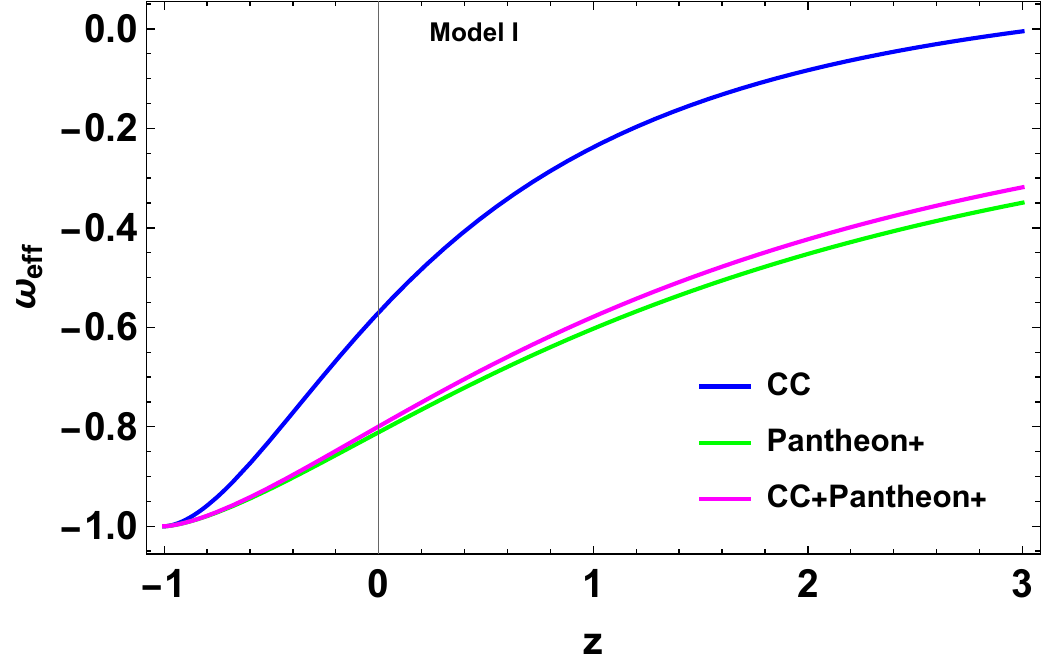}
\caption{The figure displays the behavior of the effective equation of state parameter for the MOG model I.}\label{f7a}
\end{figure}

\begin{figure}[H]
\includegraphics[scale=0.47]{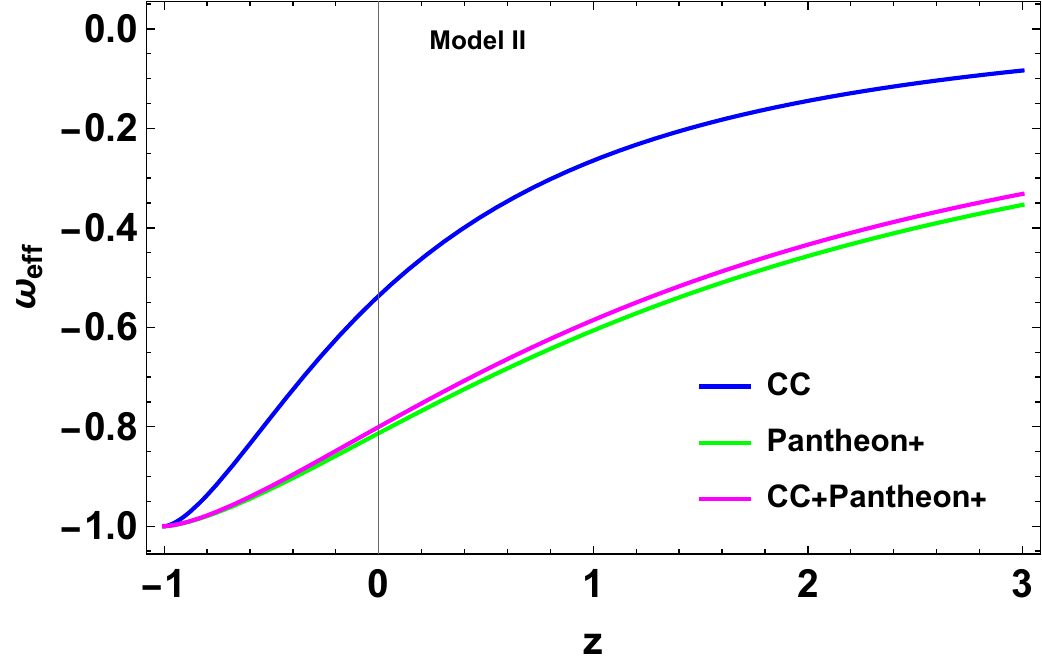}
\caption{The figure displays the behavior of the effective equation of state parameter for the MOG model II.}\label{f7b}
\end{figure}

The accelerating expansion of the universe is depicted in the behavior of the effective equation of state parameter of both considered MOG models (presented in Figs \eqref{f7a} and \eqref{f7b}). Moreover, the present time value of the EoS parameter for the model I are $\omega_0 \approx -0.569$, $\omega_0 \approx -0.810$, and $\omega_0 \approx -0.799$ and for the model II are $\omega_0 \approx -0.536$, $\omega_0 \approx -0.812$, and $\omega_0 \approx -0.8 $ corresponding to CC, Pantheon+, and the combined CC+Pantheon+ samples respectively.

The statefinder diagnostic, initially introduced by V. Sahni \cite{R37}, offers a geometric means to distinguish various dark energy models based on statefinder parameters, defined as,
\begin{equation}\label{5a}
 r =\frac{\dddot{a}}{aH^3} \:\: \text{and} \:\: s=\frac{(r-1)}{3(q-\frac{1}{2})} 
\end{equation}
The value $(r < 1, s > 0)$ indicates the quintessence dark energy, whereas the region $(r > 1, s < 0)$ signifies the phantom scenario. On the other hand, the value $(r = 1, s = 0)$ replicates the standard $\Lambda$CDM model.

\begin{figure}[H]
\includegraphics[scale=0.45]{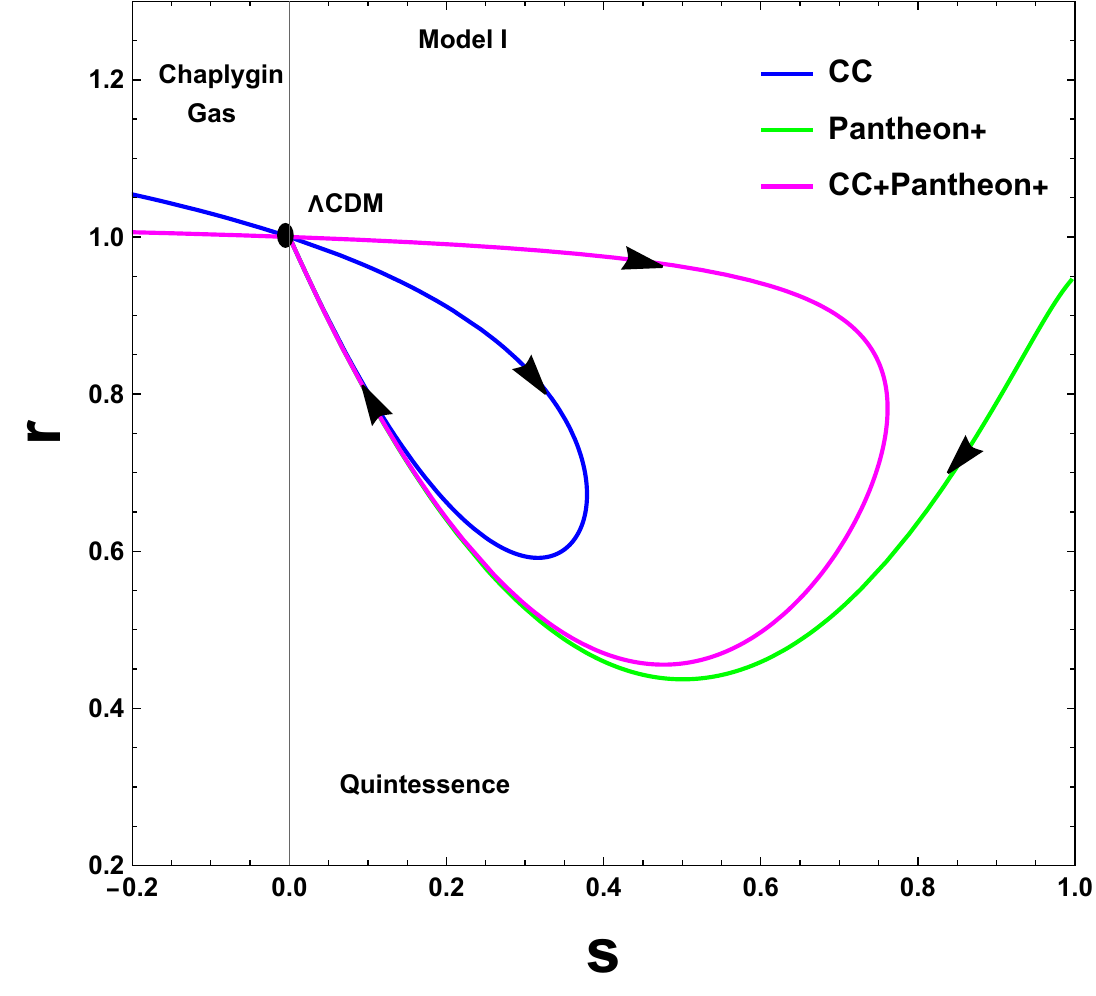}
\caption{The figure displays the behavior of the statefinder parameters for the MOG model I in the $r-s$ plane.}\label{f8a}
\end{figure}

\begin{figure}[H]
\includegraphics[scale=0.45]{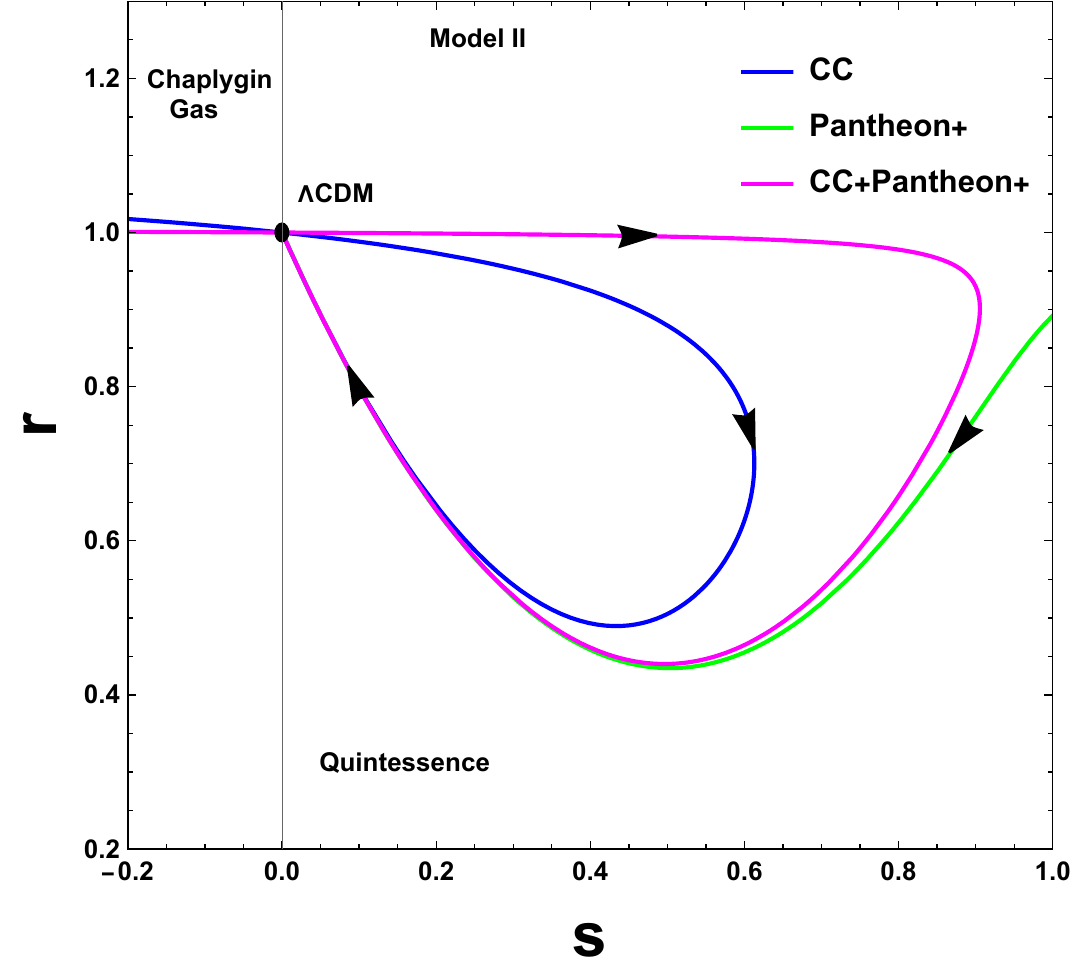}
\caption{The figure displays the behavior of the statefinder parameters for the MOG model II in the $r-s$ plane.}\label{f8b}
\end{figure}

Fig \eqref{f8a} and \eqref{f8b} illustrates the behavior of our viscous MOG models in the $r-s$ plane. The present time value of the statefinder parameters for the model I are $(r_0,s_0) = (0.59,0.31)$, $(r_0,s_0) = (0.65,0.18)$, and $(r_0,s_0) = (0.64,0.19)$ and for the model II are $(r_0,s_0) = (0.48,0.42)$, $(r_0,s_0) = (0.66,0.18)$, and $(r_0,s_0) = (0.64,0.19)$ corresponding to CC, Pantheon+, and the combined CC+Pantheon+ samples respectively. Hence, the present day values of statefinder parameters of the considered MOG models favors the quintessence type behavior.

The Om diagnostic stands as a straightforward testing method, relying solely on the first-order derivative of the cosmic scale factor. In the case of a spatially flat universe, its expression is as follows \cite{R38},
\begin{equation}\label{5b}
Om(z)= \frac{\big(\frac{H(z)}{H_0}\big)^2-1}{(1+z)^3-1}
\end{equation}
The descending slope of $Om(z)$ curve is indicative of quintessence-like behavior, whereas an ascending slope corresponds to phantom behavior. On the other hand, a constant $Om(z)$ signifies the characteristics of the $\Lambda$CDM model.

\begin{figure}[H]
\includegraphics[scale=0.48]{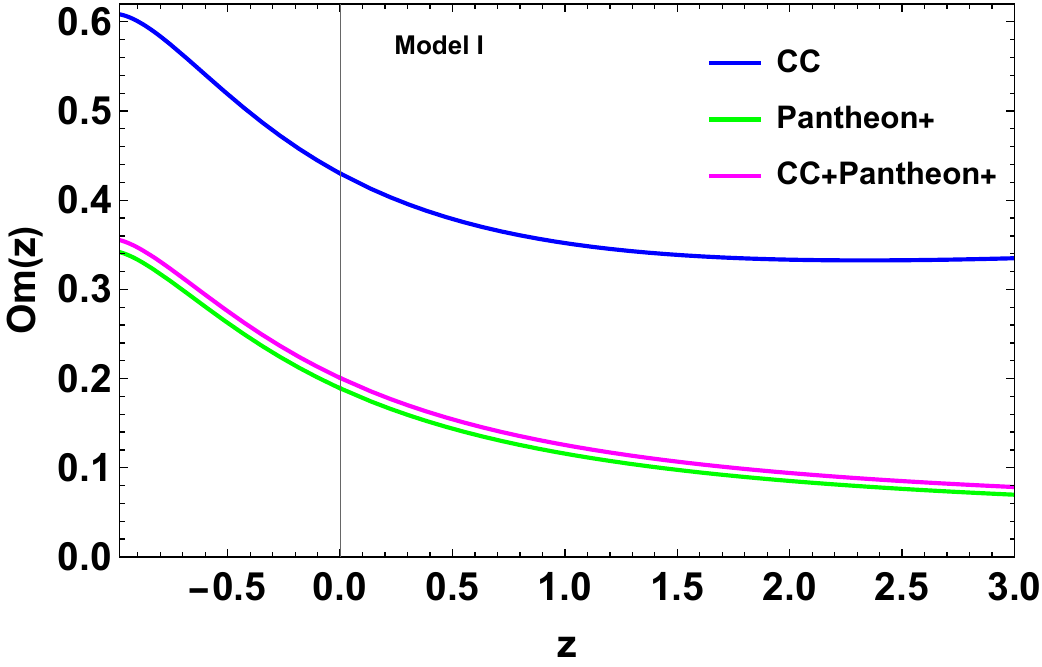}
\caption{The figure displays the behavior of the Om diagnostic parameter for the MOG model I.}\label{f9a}
\end{figure}

\begin{figure}[H]
\includegraphics[scale=0.48]{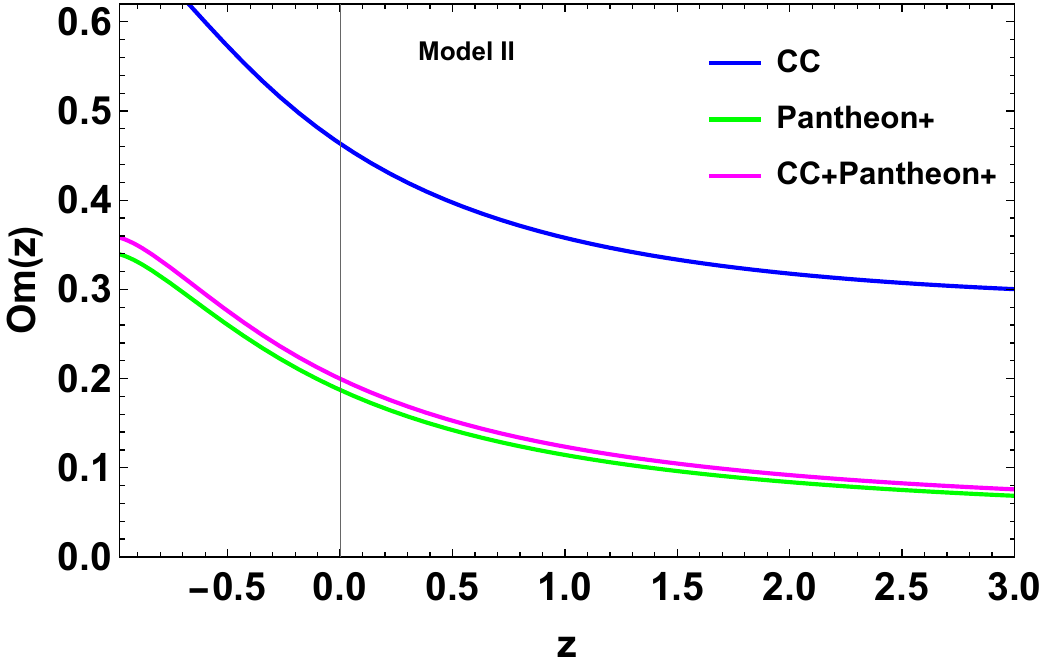}
\caption{The figure displays the behavior of the Om diagnostic parameter for the MOG model I.}\label{f9b}
\end{figure}

Fig \eqref{f9a} and \eqref{f9b} illustrates the behavior $Om(z)$ curve corresponding to considered MOG models. The presented $Om(z)$ curves have consistent negative slope across the entire domain, for all three constrained values of the model parameters. Based on this result from the Om diagnostic test, it can be inferred that our cosmological setting utilizing $f(T, \mathcal{T})$ MOG models embodies quintessence-like behavior.

\section{Conclusion} \label{sec6}
\justifying
The study of torsion-based gravity has gained substantial attention from cosmologists, encompassing a range of topics, including wormholes, black holes, late-time acceleration, inflation, etc. Meanwhile, cosmological models involving viscous fluids have gained significant interest, particularly for their description of the universe's early stages and offering insights into late-time expansion. Considering hydrodynamics, incorporating the influence of viscosity in the cosmic fluid is a reasonable assumption, given that a perfect fluid is ultimately an idealization. In this study, we explored the significance of viscosity coefficients in explaining the observed cosmic acceleration in $f(T, \mathcal{T})$ gravity background, which is an extension of the $f(T)$ gravitational theory, allowing a broad coupling between the energy-momentum scalar $\mathcal{T}$ and the torsion scalar $T$.

We begin with two $f(T,\mathcal{T})$ functions, specifically a linear model $f(T,\mathcal{T})=\alpha T + \beta \mathcal{T}$ and a non-linear model $f(T,\mathcal{T})=\alpha \sqrt{-T} + \beta \mathcal{T}$, where $\alpha$ and $\beta$ are arbitrary constants, along with the fluid part incorporating the coefficient of bulk viscosity $\zeta=\zeta_0 > 0$ adhering to a scaling law, which transforms the Einstein case into a proportional form with respect to the Hubble parameter \cite{IB-1}. The analytical solutions of the corresponding field equations for a flat FLRW environment are presented in equations \eqref{3i} and \eqref{3i1}. The free parameters of the obtained solutions have been constrained using Cosmic chronometers (CC), Pantheon+, and the CC+Pantheon+ samples. We performed the Bayesian statistical analysis to estimate the posterior probability utilizing the likelihood function and the MCMC random sampling technique. The obtained restrictions are presented in the Table \eqref{Table-2} for the Gaussian priors listed in the Table \eqref{Table-1}. In addition, the contour plots corresponding to both MOG models for the free parameters $H_0, \alpha, \beta$, and $\zeta_0$ within the $1\sigma-3\sigma$ confidence interval are presented in Figs \eqref{f2a}-\eqref{f2b}, \eqref{f4a}-\eqref{f4b}, and \eqref{f5a}-\eqref{f5b} utilizing CC, Pantheon+, and the combined CC+Pantheon+ samples respectively, along with the error bar plots of the Hubble function and the distance modulus function, presented in Figs \eqref{f1a}-\eqref{f1b} and \eqref{f3a}-\eqref{f3b}. Further, to assess the effectiveness of our MCMC analysis, we estimated the corresponding AIC and BIC values, listed in Table \eqref{Table-3}. We found that there is strong evidence supporting the assumed viscous $f(T, \mathcal{T})$ gravity models for all three considered data sets. Also, we observed that the model I is more precisely mimics the $\Lambda$CDM model. 

We also examined the evolutionary behavior of some prominent cosmological parameters. The effective equation of state parameter of both considered MOG models predicts the accelerating behavior of the expansion phase of the universe (see Figs \eqref{f7a} and \eqref{f7b}). The value of the EoS parameter at the present redshift $(z=0)$ for the model I are $\omega_0 \approx -0.569$, $\omega_0 \approx -0.810$, and $\omega_0 \approx -0.799$ and for the model II are $\omega_0 \approx -0.536$, $\omega_0 \approx -0.812$, and $\omega_0 \approx -0.8 $ corresponding to CC, Pantheon+, and the combined CC+Pantheon+ samples. Further, Figs \eqref{f8a} and \eqref{f8b} illustrate the behavior of the assumed viscous $f(T, \mathcal{T})$ models in the $r-s$ plane. The corresponding present time value of the statefinder parameters for the model I are $(r_0,s_0) = (0.59,0.31)$, $(r_0,s_0) = (0.65,0.18)$, and $(r_0,s_0) = (0.64,0.19)$ and for the model II are $(r_0,s_0) = (0.48,0.42)$, $(r_0,s_0) = (0.66,0.18)$, and $(r_0,s_0) = (0.64,0.19)$ utilizing the model parameter constraints obtained by CC, Pantheon+, and the combined CC+Pantheon+ samples respectively. Hence, the statefinder parameters of both considered MOG models favor the quintessence-type behavior. Lastly, in Figs \eqref{f9a} and \eqref{f9b}, we illustrated the behavior $Om(z)$ curve, which represents a consistent negative slope across the entire domain for both considered MOG models. Thus, it can be inferred that our cosmological setting utilizing $f(T, \mathcal{T})$ gravity models with the viscous matter fluid embodies quintessence-like behavior and can successfully describe the late-time scenario. \\

\textbf{Data availability} There are no new data associated with this article.

\acknowledgments  RS acknowledges UGC, New Delhi, India for providing Senior Research Fellowship with (UGC-Ref. No.: 191620096030). Aaqid Bhat expresses gratitude to the BITS-PILANI Hyderabad Campus, India, for granting him a Junior Research Fellowship. PKS  acknowledges the Science and Engineering Research Board, Department of Science and Technology, Government of India for financial support to carry out Research Project No.: CRG/2022/001847. 

\end{document}